\documentclass[11pt]{iopart}
\usepackage{amssymb}
\usepackage{amsbsy}
\usepackage{verbatim}
\usepackage{graphicx}
\usepackage{epstopdf}
\usepackage{color}
\usepackage{graphicx,sidecap}
\usepackage{bm}

\usepackage[colorlinks,bookmarks=false,citecolor=blue,
linkcolor=red,urlcolor=blue]{hyperref}

\catcode`@=11 
\renewcommand\tableofcontents{%
  \section*{\contentsname}%
  \@starttoc{toc}%
}
\catcode`@=12

\begin{document}

\setlength{\parindent}{0pt}

\title{Entanglement negativity and conformal field theory: 
a Monte Carlo study}

\author{Vincenzo Alba$^1$}
\address{$^1$ Department of Physics and Arnold Sommerfeld
Center for Theoretical Physics, Ludwig-Maximilians-Universit\"at
M\"unchen, D-80333 M\"unchen, Germany}

\date{\today}

\begin{abstract} 

We investigate the behavior of the moments 
 of the partially transposed reduced density matrix 
$\rho^{T_2}_A$ in critical quantum spin chains. Given
 subsystem $A$ as union of two blocks, this is 
 the (matrix) transposed  of $\rho_A$ with respect to the degrees of 
 freedom of one of the two. This is also the main ingredient for constructing the  
 {\it logarithmic negativity}.
We provide a new numerical  scheme for calculating 
efficiently all the moments of $\rho_A^{T_2}$ using  
classical Monte Carlo simulations. In particular we study    
 several combinations  of the moments 
which are {\it scale invariant} at a critical point.  
Their behavior is fully characterized in both the critical Ising 
and the anisotropic Heisenberg XXZ chains. 
For two adjacent blocks we find, in both models,  full agreement 
with recent CFT calculations. 
For  disjoint ones, in the Ising chain   
finite  size corrections  are non negligible.   
We demonstrate that their exponent is the same governing  the unusual scaling corrections of 
the mutual information between the two blocks. Monte Carlo data fully match the 
theoretical CFT prediction only in the asymptotic limit of infinite intervals.  
Oppositely, in the Heisenberg chain  scaling corrections are smaller and, 
already at finite (moderately large) block sizes,  Monte Carlo data 
are in excellent agreement with the asymptotic CFT result.
\end{abstract}

\maketitle

\section{Introduction}
\label{intro}

In recent years there has been a growing interest in 
characterizing the behavior of entanglement related
quantities (see~\cite{Renyi} for general reviews) 
in many body quantum systems. Moreover, the deep connection 
between entanglement and conformal field theory 
(CFT)~\cite{Renyi,Holzhey,cc-04,cc-rev} has boosted a huge amount of 
work at the frontiers between quantum information, condensed matter, 
and quantum field theory.

Entanglement related quantities are usually constructed considering
a bipartition of a system $S$ into two subsystems as $S=A\cup B$. 
Given the (pure) state $|\psi\rangle$ of the total system
 and the density matrix $\rho\equiv|\psi\rangle\langle\psi|$, 
the reduced density matrix for subsystem $A$ is obtained
by tracing over the degrees of freedom of $B$ as
$\rho_A\equiv \Tr_B\rho$. The  entanglement between $A$
 and $B$ can be quantified using the von Neumann
entropy $S_A\equiv-\Tr\rho_A\log\rho_A$. Alternatively, from 
 the $n-$th moment $\Tr(\rho_A^n) (n\in {\mathbb N}$)
 of the reduced density matrix  one can construct the so
called R\'enyi entropies $S_A^{(n)}\equiv1/(n-1)\log\Tr(\rho_A)^n$, 
 which are also standard entanglement measures.
  The von Neumann entropy is recovered from the R\'enyi entropies 
  as the analytic continuation $S_A=\lim_{n\to1}S_A^{(n)}$. 

\begin{figure}[t]
\begin{center}
\includegraphics[width=1.1\textwidth]{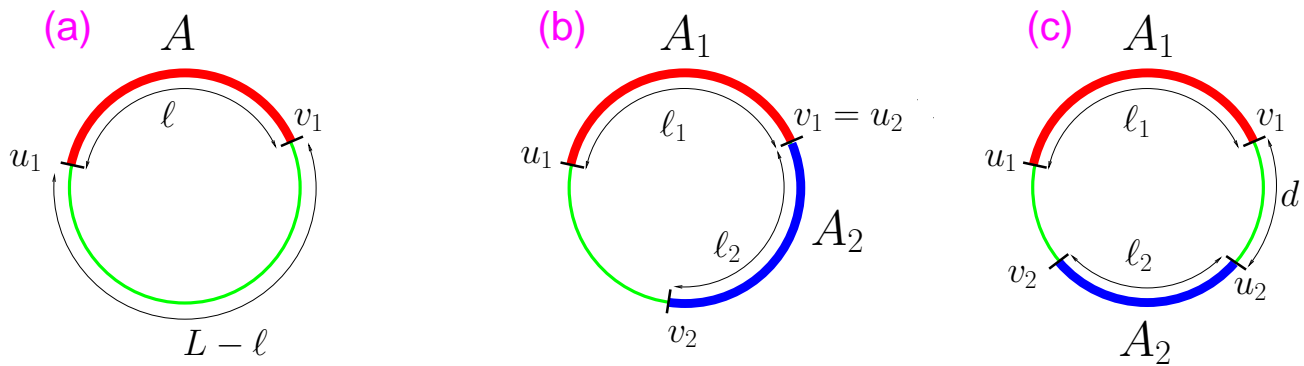}
\end{center}
\caption{ Different ways of partitioning  a 1D spin chain (of 
 length $L$) considered in this work.
 Subsystem $A$ is denoted with the thicker line. 
 ({\bf a}) $A$ is the  single interval $A\equiv [u_1,v_1]$
 (of length $\ell=|v_1-u_1|$).
  ({\bf b}) $A$ is made of two adjacent intervals 
 $A=A_1\cup A_2\equiv[u_1,v_1]\cup[u_2,v_2]$ (of length 
 respectively $\ell_1=|v_1-u_1|$ and $\ell_2=|v_2-u_2|$). 
 ({\bf c}) Two disjoint intervals $A_1,A_2$ at distance 
 $d=|u_2-v_1|$. In this work we always consider intervals of 
 equal length, i.e. $\ell_1=\ell_2=\ell$.
}
\label{cartoon_0}
\end{figure}

Let us now consider a 1D  critical quantum system (spin chain)
described in the scaling limit by a conformal field theory (CFT). 
After  restricting to periodic boundary conditions, and taking  
subsystem $A$ a single interval (Fig.~\ref{cartoon_0} ($\bf a$)), 
the scaling behavior of the R\'enyi entropies is 
given as~\cite{Holzhey,cc-04,cc-rev,Vidal,cc-05p} 

\begin{equation}
\label{ren_scal}
S_A^{(n)}=\frac{c}{6}\left(1+\frac{1}{n}\right)\log\frac{\ell}{a}
+c'_n
\end{equation} 

with $\ell$ the length of the interval.
Here $c$ is the celebrated {\it central charge}~\cite{c-lec}, $a$ an 
 ultraviolet cutoff (lattice 
spacing), and $c'_n$ a non universal constant. From~\eref{ren_scal}
 the von Neumann entropy $S_A$ is obtained as

\begin{equation}
\label{vn_scal}
S_A=\frac{c}{3}\log\frac{\ell}{a} +\tilde c_1
\end{equation}

with $\tilde c_1$ also non universal. 

Both  \eref{ren_scal}\eref{vn_scal} are nowadays accepted as  the standard tools 
to extract the central charge in 1D critical systems,  while 
other {\it universal} features can be  obtained  by analyzing their finite size 
corrections~\cite{ccen-10,ce-10,
cc-10,ccp-10,xa-11,cmv-11}.

From the field theory point of view, much more information 
about the underlying CFT is contained in the mutual information between two 
intervals~\cite{fps-08,cg-08,
cct-09,cct-11,ch-04,ffip-08,rt-06,atc-09,atc-11,f-12,ip-09,fc-10,fc-10b,c-10,rg-12}. 
 In fact, given $A$ as sum of two non complementary blocks as $A\equiv A_1\cup A_2$
 (Fig.~\ref{cartoon_0} ({\bf c})), their mutual information 
$I^{(n)}_{A_1:A_2}\equiv S^{(n)}_{A_1}+S^{(n)}_{A_2}-
S^{(n)}_{A_1\cup A_2}$  gives access to 
the full {\it operator content} of a 
CFT~\cite{tori}. From the quantum information 
perspective, however, since subsystem $A$ is not in general in a pure state, 
$I^{(n)}_{A_1:A_2}$ is {\it not} a measure of the 
mutual entanglement between  $A_1,A_2$, although it contains information
 about the correlations between them.
A standard measure of the entanglement  between two blocks 
is instead the so-called  {\it logarithmic negativity}~\cite{vw-01}.
Denoting a basis for the Hilbert space of $A_1$($A_2$) 
as $|e_i^{(1)}\rangle$ ($|e_j^{(2)}\rangle$), one first defines
 the partially transposed reduced density matrix (with respect 
 to $A_2$) $\rho^{T_2}_A$ as

\begin{equation}
\langle e_i^{(1)}\otimes e_j^{(2)}|\rho_A^{T_2}|e_k^{(1)}
\otimes e_l^{(2)}\rangle\equiv \langle e_i^{(1)}
\otimes e_l^{(2)}|\rho_A|e_k^{(1)}\otimes 
e_j^{(2)}\rangle
\end{equation}

Then the logarithmic negativity is readily given by 

\begin{equation}
\label{negativity}
{\cal E}\equiv\log||\rho_A^{T_2}||_1
\end{equation}

with $||\rho_A^{T_2}||_1$ denoting the trace norm, i.e. the 
sum of the absolute values of the eigenvalues of $\rho_A^{T_2}$.  
Besides  its importance in quantum information,
 it has been  pointed out recently that the logarithmic 
 negativity is a universal quantity at a second order phase 
 transition~\cite{Neg},  which makes ${\cal E}$ a usable 
 indicator of critical behavior in quantum many body systems.

Arguably in the context of field theories a major role is 
played by  the moments of $\rho_A^{T_2}$, i.e. $\Tr(\rho_A^{T_2})^n$.
In particular, the recent CFT approach developed in Ref.~\cite{cct-12,cct-J-12}  
provided full {\it analytical} understanding of  their scaling behavior, 
 for both adjacent and disjoint intervals (Fig.~\ref{cartoon_0} ({\bf b}) and ({\bf c})). 
In the former case  this allowed, via the analytic continuation 
$n\to 1$, to obtain an exact expression for ${\cal E}$. 
 For two disjoint intervals, although it was not possible to 
 perform the analytic continuation, it has been proven 
rigorously that at a second order phase transition 
$\cal E$ is a universal function of the (dimensionless) harmonic ratio

\begin{equation}
\label{cross_ratio}
y\equiv\frac{(v_1-u_1)(v_2-u_2)}{(u_2-u_1)(v_2-v_1)}
\end{equation}
with $u_i,v_i$ the endpoints of the two blocks (cf. Fig~\ref{cartoon_0}).
Yet,  we should mention that the  analytical treatment of  $\rho_A^{T_2}$  
is a hard task (results are available only for free 
bosons~\cite{audenaert-2002}~\cite{cct-12,cct-J-12}), and, 
as a matter of fact, a precise verification of the above mentioned CFT
findings in microscopic models  was up to now  lacking.

In this work we demonstrate that all the moments of $\rho_A^{T_2}$ 
 can be efficiently computed in classical Monte Carlo  simulations
exploiting the mapping between a quantum system in 
$d$ dimensions and a classical one in $d+1$.
The Monte Carlo technique we propose, which is itself one of our main 
 results, generalizes the one used in Ref.~\cite{cg-08,atc-09,atc-11,Gliozzi} 
to compute $\Tr\rho_A^n$ (and  the R\'enyi mutual information thereof),  
 allowing us to provide a robust verification of the CFT results 
 in Ref.~\cite{cct-12,cct-J-12}.
To this purpose, following~\cite{cct-12,cct-J-12}, 
it is convenient to define  for adjacent blocks the ratio 
$r_n$~\footnote{Note that, at difference with Ref.~\cite{cct-12}, $r_n$
 is defined here without the logarithm.} as
 
\begin{equation}
\label{rn_def}
r_n(z)\equiv\frac{\Tr(\rho_{A_1\cup A_2}^{T_2=\ell})^n}{\Tr
(\rho_{A_1\cup A_2}^{T_2=L/4})^n}
\end{equation}

with $z\equiv\ell/L$. Here the notation $\Tr\rho_A^{T_2=\ell(L/4)}$ 
means that the partial transposition is done with respect to the 
degrees of freedom of  block $A_2$ of length $\ell(L/4)$. 
For two disjoint blocks we use instead the ratio $R_n(y)$

\begin{equation}
\label{Rn_def}
R_n(y)\equiv\frac{\Tr(\rho_{A_1\cup A_2}^{T_2=\ell})^n}{\Tr
\rho_{A_1\cup A_2}^n}
\end{equation}

Remarkably both $r_n$ and 
$R_n$ are scale invariant  quantities
 at a second order transition 
(cf.~\cite{cct-12,cct-J-12} or section~\ref{summ_cft}), 
which makes them good  indicators for quantum and classical 
critical behaviors. In particular, a part from scaling corrections, 
which are expected to be less severe for ${\cal E}$~\cite{cct-12}, 
they are as effective as the logarithmic negativity. Also, their being not 
related to any local order parameter  makes them suitable especially  
for detecting {\it topological} transitions. On the CFT side, we 
anticipate here that (cf. section~\ref{summ_cft}), while the scaling 
function $r_n(z)$ is  fully characterized in terms of the central charge, 
$R_n(y)$ is a universal function of  $y$ and  depends on the full operator 
content of the given theory. In this sense  $R_n(y)$ provides yet another tool 
(besides the mutual information) to unveil the deep structure of CFTs.

\paragraph{The models.}
In this work we focus on  the critical Ising quantum spin chain and the 
anisotropic Heisenberg XXZ  model at $\Delta=-1/\sqrt{2}$ ($\Delta$ is the 
anisotropy). In both cases we consider periodic boundary conditions. 
The Ising chain in a transverse field $h$ is defined in terms 
of the Hamiltonian

\begin{equation}
{\mathcal H}^{Is}=-\sum\limits_i[\sigma_i^x\sigma_{i+1}^x+h\sigma_i^z]
\end{equation}

with $\sigma_i^{x,y,z}$ the Pauli matrices and $i=1,2,\dots,L$. The 
model exhibits a ferromagnetic (paramagnetic) phase for $h<1$ ($h>1$)
 with a second order phase transition at the critical value $h_c=1$.
 Its critical behavior is described  in the continuum by the 
free Majorana fermion theory, which is the simplest and most studied CFT 
(with $c=1/2$). The same theory describes the critical behavior
 of the 2D classical Ising model.

The  anisotropic Heisenberg XXZ spin chain  is instead defined by the 
interaction

\begin{equation}
{\mathcal H}^{XXZ} = \sum\limits_{i=1}^{L} 
(\sigma^x_i\sigma^x_{i+1}+ \sigma^y_i\sigma^y_{i+1}) ~+~
\Delta\sum\limits_{i=1}^{L} \sigma^z_i\sigma^z_{i+1}
\end{equation}
Its phase diagram  shows a  critical liquid phase for $-1<\Delta\le 1$ and  
 a gapped one at $|\Delta|>1$ (the point $\Delta=-1$ is critical but
 not conformal invariant). The liquid phase is  
  described  in the continuum limit by  the free compactified boson
theory  (or Luttinger liquid), which is  a conformal field theory 
with $c=1$~\cite{ginsparg-89,book}. The region at $-1<\Delta\le-1/\sqrt{2}$ is also mapped into 
the low-temperature critical phase of the 2D classical XY model, which
  describes a system  of  interacting {\it classical} spins (rotors) 
$\vec S_i\equiv (\cos\theta_i,\sin\theta_i)$ and is defined by the Hamiltonian

\begin{equation}
{\mathcal H}^{2D \, XY}\equiv-\beta\sum\limits_{\langle ij\rangle}\textrm{Re}[
\bar\psi_i\psi_j]=-\beta\sum\limits_{\langle ij\rangle}\vec S_i\cdot\vec S_j=
  -\beta\sum\limits_{\langle ij\rangle}\cos(\theta_i-\theta_j)
\end{equation}

Here $\psi_i\equiv e^{i\theta_i}\in U(1)$ are phases living on a 
two dimensional square lattice, $\beta=1/T$, and $\langle ij\rangle$ denotes 
nearest-neighbor sites. The XY model exhibits a low-temperature gapless critical phase 
  characterized by quasi-long-range order (QLRO). This is  divided from the standard 
paramagnetic phase at high temperature by a Berezinskii-Kosterlitz-Thouless (BKT)  
topological transition at $\beta_{BKT}=1.1199(1)$~\cite{berezinskii-71,kt-73,kt-74,
jose-78,amit-80,has-05}. The critical properties at the BKT point are the same as 
in the XXZ chain at $\Delta=-1/\sqrt{2}$, a part from logarithmic corrections that 
are present only in the classical model. 

\paragraph{Summary of the results.}
The main results of this work can be summarized as follows. 
For two adjacent blocks, in  both the critical Ising and XXZ chains and already for
 finite (large) $\ell$,  the ratio $r_n(z)$ 
 is numerically indistinguishable from its  asymptotic value 
(i.e. at $L,\ell\to\infty$), meaning that  scaling corrections are small. 
Moreover, Monte Carlo data are  in full agreement (for any value of $z$) 
with the CFT result in Ref.~\cite{cct-12}.

For disjoint intervals we focus on $R_3(y)$. For the Ising chain 
unusual (in the sense of Ref.~\cite{cc-10})  scaling corrections are 
non negligible, as observed for the mutual information~\cite{cct-11,
atc-09,atc-11,f-12,c-10} (see also~\cite{cct-12,cct-J-12}). 
We numerically demonstrate that they decay as $\ell^{-\omega_3}$ 
with $\omega_3=1/3$, in agreement with the general 
behavior (as $\ell^{-1/n}$) found in the case of   
the mutual information. 
This allows to  conclude that scaling corrections are the same 
for both quantities. By a standard finite size scaling analysis we then 
 show that the asymptotic scaling function $R_3(y)$ 
  perfectly matches  the CFT. In the Heisenberg 
chain  both usual and unusual scaling corrections are smaller, 
and already at $\ell\sim 50$   Monte Carlo data for $R_3(y)$ are in excellent 
agreement with the CFT result.

\section{Negativity and Conformal Field Theory: general results}
\label{summ_cft}

In the next  sections we briefly review the scaling behavior of $r_n(z),R_n(y)$ 
in a generic system described by conformal field theory (cf.~\cite{cct-12,cct-J-12}
 for more details). In order to make the manuscript self contained we 
 start recalling some basic facts about $\Tr\rho_A^n$ 
  and the mutual information (which  enter in the construction of $R_n(y)$) 
in section~\ref{summ_cft_renyi}. Then the behavior 
of $r_n$ and $R_n$ is discussed in section~\ref{summ_cft_ptrans}. 
Finally in~\ref{summ_cft_Rn_Is} and~\ref{summ_cft_Rn_XY} we specialize 
the result for $R_n(y)$  to the two cases of interest: the Ising universality 
class and the free compactified boson (Luttinger
liquid). The result for the Luttinger liquid has been derived already 
in~\cite{cct-J-12}, whereas the one for the Ising is derived here~\footnote{
    During the completion of this work we became aware that
 the same result has been derived by~P.~Calabrese et. al~\cite{cctt-13}} 
 using the results of Ref.~\cite{cct-12}.

\subsection{The moments of $\rho_A$ (R\'enyi entropies) \& the 
mutual information}
\label{summ_cft_renyi}

Let us consider  a 1D system described by a conformal field theory and 
 take as subsystem $A$  a single interval (as in Fig.~\ref{cartoon_0} 
({\bf a}))  of length $\ell\equiv|v_1-u_1|$.  The asymptotic scaling 
behavior of the moments $\Tr\rho_A^n$ of the reduced density matrix 
 is given as

\begin{equation}
\label{single}
\Tr\rho_A^n=c_n\ell^{-\frac{c}{6}(n-\frac{1}{n})}
\end{equation} 

with $c_n$ a non universal constant and $c$ the central 
charge of the CFT. As shown by Calabrese and Cardy in Ref.~\cite{cc-04},
 the $n$-th moment of the reduced density matrix can be also obtained in 
field theory in terms of a path integral $Z_n$ over the so-called 
$n$-sheeted Riemann surface ${\cal R}_n$ as

\begin{equation}
\label{ratio_0}
\Tr\rho_A^n=\frac{Z_n}{Z^n}
\end{equation}
where $Z$ is the same path integral (but on the plane) and ensures the correct
 normalization $\Tr\rho_A=1$. It is worth observing that~\eref{ratio_0} 
lies at the heart of all the algorithms for calculating R\'enyi entropies in both 
classical and quantum Monte Carlo simulations~\cite{cg-08,atc-09,atc-11,qmc}  
(as it will be better clarified  in section~\ref{MC_procedure}).

On the field theory  side one further notices that~\eref{ratio_0} can be 
rewritten in terms of the so-called  branch point {\it twist fields} 
${\cal T}$ (and anti-twist $\bar{\cal T}$) as 

\begin{equation}
\label{twist_single}
\textrm{Tr}\rho_A^n=\langle{\cal T}_n(u_1)\bar {\cal T}_n(v_1)
\rangle
\end{equation}

The twist(and anti-twist) fields are primary fields (in the CFT language) 
and are inserted respectively at the endpoints $u_1$ and $v_1$ of interval $A$
 (Fig.~\ref{cartoon_0}). Their scaling dimensions $\Delta_n=\bar\Delta_n$ are given as 

\begin{equation}
\label{twist_dim}
\Delta_n=\frac{c}{12}\Big( n-\frac{1}{n}\Big)
\end{equation}

Using~\eref{twist_dim} and basic properties of correlation functions, 
  it is a simple exercise in CFT to obtain~\eref{single} from~\eref{twist_single}.

For two disjoint intervals (see Fig.~\ref{cartoon_0} ($\bf c$)) the moments 
$\Tr\rho_{A_1\cup A_2}^n$ admit  a similar representation 
in terms of twist fields  and one now obtains the four point function

\begin{equation}
\label{trA_twist}
\Tr\rho_{A_1\cup A_2}^n=\langle{\cal T}_n(u_1)\bar{\cal T}_n
(v_1){\cal T}_n(u_2)\bar{\cal T}_n(v_2)\rangle
\end{equation}

which in {\it any} CFT, using only the global conformal invariance,  
can be  recast as 

\begin{equation}
\label{trA}
\Tr\rho_{A_1\cup A_2}^n=c_n^2\big[\ell_1\ell_2(1-y)\big]^{-\frac{c}{6}
(n-\frac{1}{n})}{\cal F}_n(y)
\end{equation}

Here  $y$ is the harmonic ratio~\eref{cross_ratio} and  ${\cal F}_n(y)$ 
(for each $n$) a universal scaling function   
containing complete information about the underlying  CFT. 
For example, the  Taylor expansion of ${\cal F}_n(y)$ at small $y$ is also 
universal and allows to extract the full operator content (scaling 
dimensions, OPE (operator product expansion) coefficients, etc.) 
of the theory~\cite{cct-09,cct-11}. From~\eref{trA} the scaling behavior of 
the R\'enyi mutual information $I^{(n)}_{A_1:A_2}$ is given as

\begin{equation}
\label{mi}
 I^{(n)}_{A_1:A_2}\equiv\log\frac{\Tr\rho^n_{A_1\cup A_2}}
{\Tr\rho^n_{A_1}\Tr\rho_{A_2}^n} =\log\Big[(1-y)^{-\frac{c}{6}
(n-\frac{1}{n})}{\cal F}_n(y)\Big]
\end{equation}

In constructing~\eref{mi} the non universal factors $c_n$ 
appearing in~\eref{trA} cancel, implying  that the R\'enyi mutual information 
$I_{A_1:A_2}^{(n)}$ is a universal function of solely the  
harmonic ratio $y$. A part from the  ``trivial'' factor $(1-y)^{-c(n-1/n)/6}$ it depends only 
on the function ${\mathcal F}_n(y)$.  A notable consequence is that the mutual 
 information allows to distinguish between CFTs with the same central charge
 but different operator content. The most prominent example is perhaps the 
 Luttinger liquid, whose operator content changes 
 as a function of the Luttinger parameter $K_L$, although one has
 $c=1$ independently of $K_L$.

On the other hand, one should mention that exact results  
for ${\cal F}_n(y)$ are known so far only for few CFTs, 
namely  the free compactified boson theory~\cite{cct-09} and the Ising 
universality class~\cite{cct-11}. Also, even for the aforementioned 
models, performing the analytic continuation $n\to 1$ to get the von 
Neumann mutual information $I_{A_1:A_2}$ represents  still a formidable 
task and results are only available  in some limits~\cite{cct-09,cct-11}.
One reason why it is  desirable to calculate $I_{A_1:A_2}$ is that, 
while $I_{A_1:A_2}^{(n)}$ exhibit strong unusual corrections  
(often oscillating), making any attempt to extract ${\cal F}_n(y)$ in 
microscopic models numerically demanding, for 
the von Neumann mutual information $I_{A_1:A_2}$ scaling  corrections are 
usually smaller~\cite{fps-08,atc-09,atc-11,f-12,fc-10,fc-10b}.

\subsection{The moments of $\rho_A^{T_2}$, logarithmic negativity, 
and the ratios $r_n$ $R_n$}
\label{summ_cft_ptrans}

In this section we discuss  the behavior of the moments  of the 
partially transposed reduced density matrix   $\rho_{A_1\cup A_2}^{T_2}$ 
and the logarithmic negativity $\cal E$ in systems described by CFTs. 
It has been observed in Ref.~\cite{cct-12} that 
$\Tr(\rho_{A_1\cup A_2}^{T_2})^n$  can be written 
in terms of the same  twist fields appearing in~\eref{twist_single}. In 
the most general  case of two disjoint blocks $A_1,A_2$ one has

\begin{equation}
\label{trA_tran_twist}
\Tr(\rho_{A_1\cup A_2}^{T_2})^n=\langle{\mathcal T}_n(u_1)
\bar{\mathcal T}_n(v_1)\bar{\mathcal T}_n(u_2)
{\mathcal T}_n(v_2)\rangle
\end{equation}

while the case of two adjacent ones can be  recovered as the limit $u_2\to v_1$
 (cf. Fig~\ref{cartoon_0}).
One should notice that \eref{trA_tran_twist} can be  obtained from~\eref{trA_twist} 
  by replacing ${\cal T}(u_2)\to \bar{\cal T}(u_2)$ and
 $\bar{\cal T}(v_2)\to {\cal T}(v_2)$. 
 This could be seen,  somehow, as the implementation in the field theory 
language of the partial transposition~\footnote{
There is a subtlety here: formula~\eref{trA_tran_twist} would correspond to 
$C\rho_A^{T_2}C$ (not $\rho_A^{T_2}$) with $C$ the transformation
 reversing the order of rows and columns indices referring to the second interval
 $A_2$. However, the transformation $C$ does not affect the moments of 
$\rho_A^{T_2}$~\cite{cct-12,cct-J-12}.}. From~\eref{trA_tran_twist}, 
using only the global conformal symmetry, the scaling 
behavior of $\Tr(\rho_{A_1\cup A_2}^{T_2})^n$ 
can be obtained as follows~\cite{cct-12,cct-J-12}.

\subsubsection{Adjacent intervals.}
We first discuss the case of two adjacent blocks, 
i.e. at distance $d=0$  (see Fig.~\ref{cartoon_0}
 ($\bf b$)). Then it can be shown that~\eref{trA_tran_twist} reduces to 
 a three twists correlation function, whose form is fully 
determined by the conformal symmetry. The result depends only on 
 the central charge and, surprisingly,  
on the parity of $n$. Its exact form is  given as~\cite{cct-12,cct-J-12}

\begin{equation}
\label{rn_cft}
\Tr(\rho_{A_1\cup A_2}^{T_2})^n\propto\left\{
\begin{array}{cc}
(\ell_1\ell_2)^{-\frac{c}{6}(\frac{n}{2}-\frac{2}{n})} 
(\ell_1+\ell_2)^{-\frac{c}{6}(\frac{n}{2}+\frac{1}{n})}
& n \,\textrm{even}\\
\\
(\ell_1\ell_2(\ell_1+\ell_2))^{-\frac{c}{12}(n-\frac{1}{n})} 
& n \,\textrm{odd}
\end{array}\right.
\end{equation}

with $\ell_1(\ell_2)$  the size of $A_1(A_2)$ ($\ell_i$ given as
 $\ell_i=|v_i-u_i|$). We remark that~\eref{rn_cft} holds provided 
that $1\ll\ell_1,\ell_2\ll L$ (i.e. for two intervals embedded in 
an infinite chain), while for finite size chains one should replace, as 
usual in CFT, $\ell_i\to L/\pi\sin(\pi\ell_i/L)$ (chord length). The exact  
functional form of $r_n(z)$ now can be obtained from~\eref{rn_cft} and~\eref{rn_def}. 
 Since  it is quite clumsy, we do not report the explicit result, which, 
instead, will be shown numerically in Fig.~\ref{fig1} and Fig.~\ref{fig3_xy} for   
 respectively $c=1/2$ and $c=1$ ($n=3,4$).

Finally, the logarithmic negativity  $\cal E$  for two adjacent 
blocks is obtained performing the analytic continuation  
$n\to 1$
 of $\log\Tr(\rho_A^{T_2})^n$  (only using $n$ 
even in~\eref{rn_cft}~\footnote{The analytic continuation for $n$ odd gives the 
normalization condition $\Tr\rho_A^{T_2}=1$~\cite{cct-12,cct-J-12}.}). The result 
can be given as

\begin{equation}
{\cal E}= \frac{c}{4}\log\frac{\ell_1\ell_2}{\ell_1+\ell_2}+\textrm{cnst.}
\end{equation}

and is universal~\cite{cct-12}.

\subsubsection{Disjoint intervals.}
In the (more complex) case of  $A$ being made of two disjoint 
intervals the scaling behavior of  $(\Tr\rho_A^{T_2})^n$   is given as

\begin{equation}
\label{trAt}
\Tr(\rho_{A_1\cup A_2}^{T_2})^n=c_n^2[\ell_1\ell_2(1-y)]^
{-\frac{c}{6}(n-\frac{1}{n})}{\cal G}_n(y)
\end{equation}

with ${\cal G}_n(y)$ a universal function of the harmonic 
ratio $y$. As for the mutual information (cf. previous section), the form 
of~\eref{trAt} is fixed only by the global conformal invariance 
of~\eref{trA_tran_twist}. Remarkably ${\cal G}_n(y)$ can 
be related to the scaling function ${\cal F}_n(y)$ appearing 
in~\eref{trA} as~\cite{cct-12,cct-J-12}

\begin{equation}
\label{Rn_cft}
{\cal G}_n(y)=(1-y)^{\frac{c}{3}(n-\frac{1}{n})}
{\cal F}_n(y/(y-1))
\end{equation}

In  constructing the ratio $R_n(y)$ (cf.~\eref{Rn_def}) 
 all the  non universal factors in~\eref{trAt} cancel and one obtains
 a universal function of the harmonic ratio

\begin{equation}
\label{cft_result}
R_n(y)=(1-y)^{\frac{c}{3}(n-\frac{1}{n})}
\frac{{\cal F}_n(y/(y-1))}{{\cal F}_n(y)}
\end{equation}
 
It is interesting to investigate the asymptotic behavior of $R_n(y)$
 in the limits $y\to 0,1$. At  $y\to 1$ one should recover the result
 for two adjacent intervals: according to~\eref{cross_ratio}, in fact, 
$y\to 1$  corresponds to  $u_2\to v_1$. Comparing~\eref{Rn_cft} with
~\eref{rn_cft} one obtains that $\Tr(\rho_{A_1\cup A_2}^{T_2})^n
\sim (1-y)^\gamma$ with $\gamma=c(n-1/n)/12$ if $n$ is odd and $\gamma=c(n/2-2/n)/6$ 
for $n$ even. As a consequence the asymptotic behavior of $R_n(y)$,
 which depends on the central charge and on the parity of $n$, 
is given as

\begin{equation}
\label{asy_y1}
R_n(y)\sim\left\{\begin{array}{cc}
(1-y)^{\frac{c}{12}(n-\frac{1}{n})} & n\quad \textrm{odd}\\\\
(1-y)^{\frac{c}{6}(\frac{n}{2}-\frac{2}{n})} & n\quad \textrm{even}
\end{array}\right.
\end{equation}

On the contrary, the behavior 
in the limit $y\to 0$ can be obtained using the methods reported in 
Ref.~\cite{cct-11} and is expected to depend in general on the operator 
content of the CFT.
 
To conclude we mention that, since ${\cal G}_n(y)$ shows in general a non trivial
 dependence on $n$, it is tricky to perform the analytic continuation 
$n\to 1$ to obtain ${\cal E}$. Despite that, according to~\eref{trAt} one has that,  
 formally, ${\cal E}$ is given as $\lim_{n\to 1}\log({\cal G}_n(y))$ (only using
even $n$, as for two adjacent blocks). This allows to conclude that the logarithmic
 negativity is a universal function of $y$, as already argued in Ref.~\cite{Neg}
 on the basis of DMRG data. Furthermore,  for some CFTs exact results for ${\cal E}$ 
have been worked out in the limit $y\to 1$~\cite{cct-12,cct-J-12}. 
A surprising result, from the CFT perspective, is that  
in the limit $y\to 0$ the logarithmic negativity apparently is vanishing  
faster than any power (i.e. in  a non analytic way)~\cite{cct-12,cct-J-12},
 whereas $R_n(y)$, for {\it any} $n$, is analytic  
 in $y=0$.

\section{The universal ratio $R_n(y)$ in the 1D Ising 
universality class}
\label{summ_cft_Rn_Is}

In this section we derive the  universal scaling function $R_n(y)$ 
in the Ising universality class. To this purpose we remind that 
for the Ising model  the scaling function ${\cal F}_n(y)$ appearing 
in~\eref{cft_result} is given as~\cite{cct-11}

\begin{equation}
{\cal F}_n (x)= 
\frac{1}{2^{n-1}\Theta({\bf 0}|\Gamma)} \sum_{
\bm{\varepsilon,\delta}}\Big| \Theta\bigg[\begin{array}{c} 
\bm{\varepsilon} \\ \bm{\delta}  \end{array}\bigg] 
({\bf 0}|\Gamma)\Big|
\label{fn_is}
\end{equation}

where $\Theta$ is the Riemann theta function with characteristic 
defined as

\begin{equation}\fl
\label{ri_theta}
\Theta\bigg[\begin{array}{c} \bm{\varepsilon} 
\\ \bm{\delta}  \end{array}\bigg] ({\bf z} |\Gamma)
\,\equiv\,\sum_{{\bf m} \in \mathbb{Z}^{n-1}} \exp\Big[i \pi 
({\bf m+{\bm \varepsilon}})^{{\rm T}} \,\Gamma\,
({\bf m+\bm{\varepsilon}})+2\pi i\,({\bf m+\bm{\varepsilon}}
)^{{\rm T}} ({\bf z+\bm{\delta}})\Big]
\end{equation}

with  $\bf z,{\boldsymbol\varepsilon},{\boldsymbol\delta}$ 
 vectors in ${\mathbb C}^{n-1}$. Precisely, in~\eref{fn_is}
the sum is over all the possible $n$ dimensional vectors 
${\boldsymbol\varepsilon},{\boldsymbol\delta}$ with entries 
$0,1/2$. The $(n-1)\times (n-1)$ matrix $\Gamma$ is defined as  

\begin{equation}
\label{gamma}
\Gamma_{rs} =\frac{2i}{n} \sum_{k\,=\,1}^{n-1}
\sin\left(\pi\frac{k}{n}\right)\beta_{k/n}\cos
\left[2\pi\frac{k}{n}(r-s)\right]
\end{equation}

Here  $\beta_q(x)$ is given by 

\begin{equation}
\beta_q=\frac{\, _2 F_1(q,1-q;1;1-x)}{\, _2 F_1(q,1-q;1;x)}
\end{equation}

\begin{figure}[t]
\begin{center}
\includegraphics[width=.7\textwidth]{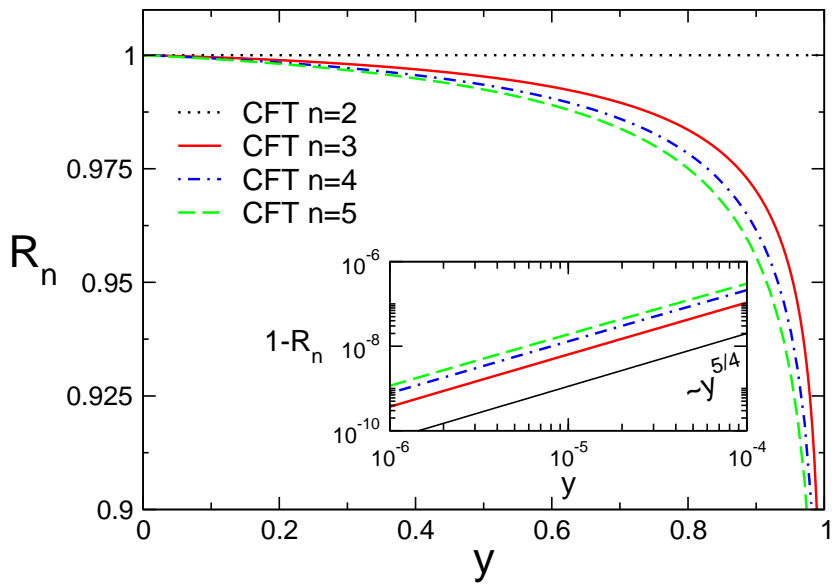}
\end{center}
\caption{ Scaling function $R_n(y)$ (CFT prediction) 
 for the 1D Ising universality class. We show $R_n(y)$ 
 as a function of the harmonic ratio $y$ for $n=2,3,4,5$. The 
 inset is to highlight the small $y$ behavior of $R_n(y)$. 
 The continuous black line is $\sim y^{5/4}$.
}
\label{fig0_is}
\end{figure}

and  $_2F_1$ is the hypergeometric function. 
Notice that for $n=2$~\eref{fn_is} can be expressed in terms
of elementary functions as~\cite{atc-09} 

\begin{equation}
\fl
{\cal F}_2(x)=\frac1{\sqrt{2}}\Bigg[\left(\frac{(1 + 
\sqrt{x}) (1 + \sqrt{1 - x})}2\right)^{1/2} + x^{1/4} 
+ ((1 - x) x)^{1/4} + (1 - x)^{1/4} \Bigg]^{1/2}
\label{CFTF2}
\end{equation}

The exact analytic form of $R_n(y)$ in the Ising universality class 
is obtained  from~\eref{fn_is} and~\eref{cft_result}.  
This is shown numerically in Fig.~\ref{fig0_is} ($R_n(y)$  as a 
function of the harmonic ratio $y$ for $n=2,3,4,5$). Clearly 
$R_n(y)$ (for any $n$) is very close to one in the 
region $y\sim 0$  and is  monotonically vanishing
 in the  limit $y\to 1$. Note that $R_2(y)$ is 
exactly one ($R_2=1$ $\forall y$). The asymptotic behavior of $R_n(y)$ in the
limit $y\to 1$,  when the two intervals are next to each other (cf. 
Fig.~\ref{cartoon_0}), is given by~\eref{asy_y1}. For instance, for $n=3,4$ one has 
$R_3(y)\sim (1-y)^{1/9}$ and $R_4(y)\sim (1-y)^{1/8}$, which explains 
the slowly vanishing behavior observed in Figure~\ref{fig0_is}.

Oppositely, in the limit $y\to 0$, when the two intervals are very far apart 
($d\gg 1$ in Fig.~\ref{cartoon_0}), the same asymptotic behavior, 
  $R_n(y)\sim 1-\alpha_n y^{5/4}$, for all the values of $n>2$ is observed 
 (this is highlighted in Fig.~\ref{fig0_is}: Inset). In principle the 
exponent $5/4$ could be calculated analytically using the same methods employed 
in Ref.~\cite{cct-11}  to obtain the small $y$ expansion of ${\cal F}_n(y)$. Note
 that one should expect  $\alpha_n\to 0$ in the limit $n\to 1$, reflecting 
 that the negativity vanishes faster than any power at $y\to 0$~\cite{cct-12,cct-J-12}.

\section{The universal ratio $R_n$ in the free compactified boson theory}
\label{summ_cft_Rn_XY}

In this section we re-derive (for more details cf.~\cite{cct-J-12}) 
the asymptotic scaling function $R_n(y)$ for a free compactified boson
 theory. This is defined by the field theory action

\begin{equation}
\label{fcb}
S=\frac{1}{2\pi}\int dzd\bar z \partial\phi\bar\partial\phi
\end{equation}

where $\phi$ are bosonic fields compactified on a circle of radius 
$r_{circle}$, meaning that $\phi=\phi+2\pi r_{circle}$. 
The action~\eref{fcb}  is also  the so-called Luttinger liquid, 
which is a $c=1$ CFT and one of the most successful paradigms
 to understand the physics of 1D systems. For instance the  theory describes
 the critical long wavelength behavior of the anisotropic Heisenberg 
spin chain in its gapless phase, bosons with repulsive delta 
interaction, 1D Hubbard model, etc.. 

For the free compactified boson theory~\eref{fcb} 
the scaling function ${\cal F}_n(x)$ has been calculated 
in Ref.~\cite{cct-09}  and is given as

\begin{equation}
\label{eq_78}
\mathcal{F}_n(x)=
\frac{\Theta\big({\bf 0}|\eta\Gamma\big)\,\Theta\big({\bf 0}
|\Gamma/\eta\big)}{[\Theta\big({\bf 0}|\Gamma\big)]^2}
\label{fn_XY}
\end{equation}

where $\Theta$ and $\Gamma$ are the same as for the 
Ising universality class (cf. previous section). Here 
$\eta$ is  related to the compactification radius $r_{circle}$ 
as $\eta=2r^2_{circle}$ and can be also given in  terms of the 
so-called Luttinger liquid parameter $K_L$ as $\eta=1/(2K_L)$. 
For $n=2$~\eref{fn_XY} is expressed in terms of the Jacobi theta 
functions $\theta_\nu$ as~\cite{fps-08}

\begin{equation}
\label{f2boson}
{\cal F}_2(y)=\frac{\theta_3(\eta\tau)\theta_3(\tau/\eta)}{
[\theta_3(\tau)]^2}
\end{equation}

where $\tau$ is related to the harmonic ratio as 
$y=[\theta_2(\tau)/\theta_3(\tau)]^4$. Notice in 
both~\eref{eq_78}\eref{f2boson} the explicit invariance under 
$\eta\to1/\eta$. The point   $\eta=4$ ($r_{circle}=\sqrt{2}$) 
describes  the critical behavior of the 2D  XY model 
at the BKT transition (or equivalently the long wavelength properties of  
the 1D Heisenberg XXZ chain at anisotropy $\Delta=-1/\sqrt{2}$) 
(cf.~\cite{ginsparg-89}).

\begin{figure}
\begin{center}
\includegraphics[width=.85\textwidth]{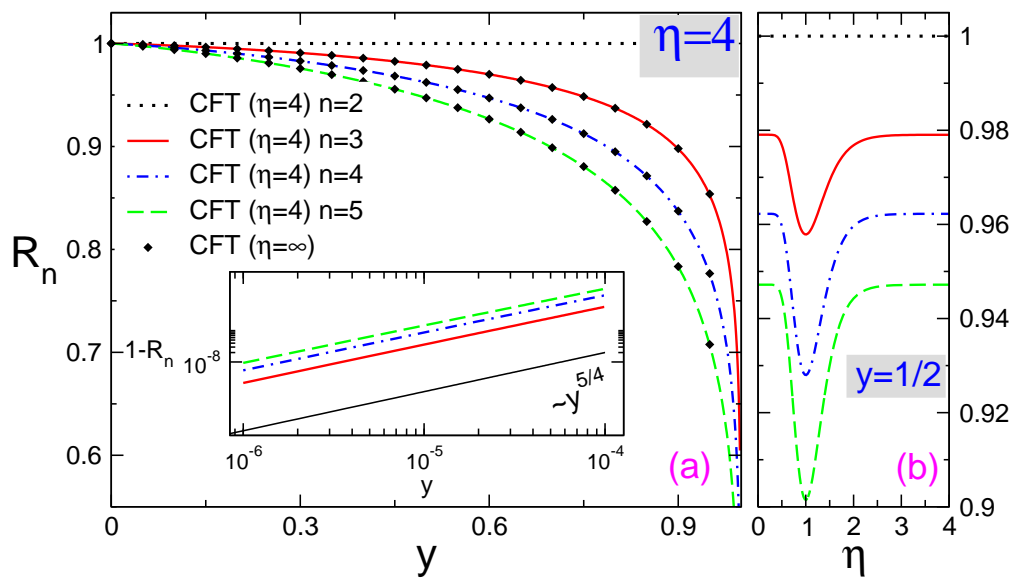}
\end{center}
\caption{ ({\bf a}) Scaling function $R_n(y)$ for the free 
 compactified boson theory  at $\eta=4$ ($r_{circle}=\sqrt{2}$). 
 CFT prediction versus the harmonic ratio $y$ for $n=2,3,4,5$. 
 For each $n$  the CFT result for the free boson in the 
 decompactified limit (i.e. at $\eta=\infty$) is also shown (rhombi). 
 Inset: small $y$ behavior of $R_n(y)$, plot of $1-R_n(y)$ versus
 $y$. The continuous black line is $\sim y^{5/4}$.
 ({\bf b}) $R_n(y)$ ($n=3,4,5$) at fixed $y=1/2$ as a function of $\eta=
 2r^2_{circle}$.
}
\label{fig0_xy}
\end{figure}

Before proceeding one should remark that~\eref{eq_78} is only defined 
for positive values of its argument $x$. Since in~\eref{cft_result} 
the combination $y/(y-1)$ is negative for $y\in [0,1)$, 
one should consider the analytic continuation for negative argument 
of~\eref{eq_78}. This has been calculated in Ref.~\cite{cct-J-12} and
 is given as

\begin{equation}
\label{fn_free_c}
{\mathcal F}_n(y)=
\frac{\eta^\frac{n-1}{2}\Theta\Big({\bf 0}|\eta\, G\big[y/(y-1)\big]
\Big)}{\sqrt{\prod\limits_{k=1}^{n-1}
\textrm{Re}(F_{k/n}(\frac{y}{y-1})\bar F_{k/n}(
\frac{1}{1-y}))}}
\end{equation}

where $F_q(x)\equiv _2F_1(q,1-q,1,x)$ and we defined $G$  
as

\begin{equation}
G\equiv2i\left(\begin{array}{cc}
A & W\\
W^T & B
\end{array}\right)
\end{equation}

The  matrix elements  of $G$ are given as 

\begin{eqnarray}
\fl
A=\sum\limits_{k=1}^{n-1}\frac{|\tau_{k/n}|^2}
{\beta_{k/n}}\sin(\pi k/n)E_{k/n}  \qquad  
B=\sum\limits_{k=1}^{n-1}\frac{1}
{\beta_{k/n}}\sin(\pi k/n)E_{k/n} & \\
\nonumber W=-\sum\limits_{k=1}^{n-1}\frac{\alpha_{k/n}}
{\beta_{k/n}}\sin(\pi k/n)[\sin(\pi k/n)-i\cos
(\pi k/n)]E_{k/n} & 
\end{eqnarray}

where we used the definitions

\begin{equation}
\tau_{k/n}= i\frac{F_{k/n}(1-x)}{F_{k/n}(x)}
\equiv\alpha_{k/n}+i\beta_{k/n}, \qquad 
(E_{k/n})_{rs}\equiv e^{2\pi i k/n(r-s)}/n
\end{equation}

The ratio  $R_n(y)$ can now  be obtained using~\eref{fn_free_c}
 and \eref{cft_result}. This is shown numerically in Fig.~\ref{fig0_xy} ({\bf a}) 
 plotting  $R_n(y)$ as a function of $y$ for $n=2,3,4,5$. 
The form of $R_n(y)$ is very similar to the Ising case 
(Fig.~\ref{fig0_is}). Precisely, $R_n(y)\approx 1$ in the whole
region $0\le y\lesssim 1$, while in the limit $y\to1$ $R_n(y)$ vanishes  
 (faster than in the Ising case 
 due to the larger value of the central charge, cf.~\eref{asy_y1}). 
 Note, however, that $R_n(y)$ ranges in a larger interval compared to the
  Ising case: for instance it is $0.9\lesssim R_3(y)\le 1$ for $0\le y\lesssim 3/4$, 
  while in the same region one has $0.98\lesssim R_3(y)\le 1$ for the Ising 
 (cf. Fig.~\ref{fig0_is}).
  
 Remarkably, the same asymptotic behavior $R_n(y)\sim 1-\alpha'_n y^{5/4}$, 
 as in the Ising case, is shown in the limit $y\to 0$ (Fig.~\ref{fig0_xy}: Inset). 
 Also  it should be $\alpha'_n\to 0$ at $n\to 1$ (as proven 
 analytically in~\cite{cct-J-12} for arbitrary values of the compactification radius).

Interestingly,  $R_n(y)$ depends weakly on the compactification 
radius $r_{circle}$, hence on $\eta$. For example in the range $2\lesssim\eta\le\infty$, 
 with $\eta=\infty$ the so-called decompactified limit, 
$R_n(y)$ does not change significantly as a function of $y$ 
 (in  Fig.~\ref{fig0_xy} the  difference between  $\eta=\infty$ 
 and $\eta=4$ is not visible at all). This is better highlighted 
 in Fig.~\ref{fig0_xy} ({\bf b})  showing $R_n(y)$ at fixed $y=1/2$ as a function 
 of $\eta=2r^2_{circle}$. Notice that, since $R_n(y)$ inherits the symmetry under 
 $\eta\to 1/\eta$ from the ${\cal F}_n(y)$ (cf.~\eref{eq_78}), one has $R_n(\eta)=R_n(1/\eta)\,
\forall (y,n)$.
 In Fig.~\ref{fig0_xy} for each $n$ $R_n(y)$ exhibits a minimum at $\eta=1$,
 which corresponds to the antiferromagnetic Heisenberg XXX model, 
 while in the region $\eta>1$ it  increases monotonically showing a saturating 
behavior for large $\eta$. In particular already for $\eta\gtrsim 2$ $R_n(y)$ cannot be 
distinguished  from its asymptotic value. The same weak dependence on $\eta$ is observed 
 at other values of $y$. One must mention that it is possible to calculate
 $R_n(y)$ directly in the decompactified limit (i.e. $\eta\to\infty$) 
 using~\eref{fn_free_c} (cf. Ref.~\cite{cct-12} for the analytic result). 

\section{The moments of $\rho_{A}^{T_2}$ in Monte Carlo simulations}
\label{MC_procedure}

In this section we present a Monte Carlo scheme for
calculating $\Tr(\rho_{A_1\cup A_2}^{T_2})^n$ using the replica trick
and classical Monte Carlo simulations.  This is based 
on the approach developed in~\cite{cg-08,atc-09,atc-11} for  
$\Tr\rho_A^n$ (R\'enyi entropies) and can be in principle applied to any 
model that can be simulated with classical Monte Carlo.
The method employs the mapping between a quantum system in $d$
 dimensions and a classical one living in $d+1$. 
We should mention that an alternative numerical approach for calculating
 both $\Tr(\rho_A^{T_2})^n$ and ${\cal E}$ using TTN 
(tree tensor network) techniques~\cite{mps,TTN,fnw-92,f-97,hieida,lcp-00,
mrs-02,sdv-06,nagaj-08,silvi,Dur,Murg,Plenio}  has 
become available recently~\cite{cctt-13}. 

The section is organized 
 as follows. We first review in~\ref{replica_trick} the replica representation 
for the moments of both $\rho_A$ and $\rho_A^{T_2}$, which lies at the heart 
 of the method. The moments  can be measured 
in Monte Carlo simulations using the strategy outlined in~\ref{MC_observable}.
Finally, in~\ref{MC_proc_clusters} we provide an improved scheme  for models 
admitting a representation in terms of cluster variables.

\subsection{Replica trick}
\label{replica_trick}

The partition function $Z=\Tr e^{-\beta H}$ of a $d$-dimensional 
 quantum system (defined in terms of an Hamiltonian $H$) 
at inverse temperature $\beta$ can be written as an Euclidean path 
integral in $d+1$ dimensions as 

\begin{equation}
Z=\int{\cal D}[\phi]e^{-S(\{\phi\})}
\end{equation}
where $\phi (\vec x,\tau)$ is a field living on the hypercubic lattice 
$\{\vec x,\tau\}$ and $S$ the Euclidean action.  The spatial coordinates 
$x_i$ are such that $0\le x_i<L_i$ with $i=1,2,\dots,d$ and the imaginary 
time $\tau$ ranges in the interval $0\le \tau<L_\tau=\beta$. 
The fields $\phi$ are periodic along the imaginary 
time direction, i.e. $\phi(\vec x,\tau+\beta)=\phi(\vec x,\tau)$.

 Here we consider the $n-$th ($n\in {\mathbb N}$)
power of the partition function (or the replicated partition function) 
which reads

\begin{equation}
Z^n=\int\prod\limits_{k=1}^n{\cal D}[\phi_k]e^{-\sum
\limits_{k=1}^n S(\{\phi_k\})}
\end{equation}

where $\phi_k\equiv\phi_k(\vec x, \tau)$ is now a field living on the 
$k$-th replica and $S(\phi_k)$ is the replica Euclidean action. 
The actual form of the action $S$ is not important for the 
following, but for the sake of simplicity we restrict to the case of  
nearest-neighbor interactions (which include the models treated in this
work) and we consider $1+1$ dimensions.  Thus we assume that 
the  action $S$ (defined on the $k-$th replica) is of the form

\begin{equation}
\label{act}
S(\phi_k)=\sum\limits_{\langle ij\rangle}
F(\phi_k(i),\phi_k(j))
\end{equation}

where $\langle ij\rangle$ denotes nearest-neighbor sites and the 
function $F$ models the interaction between the fields $\phi$. 
Since we consider periodic spin chains (see Fig.~\ref{cartoon_0}) 
we assume on each replica periodic boundary conditions also along the 
spatial direction.

\subsubsection{Replica representation for the moments of $\rho_A$ 
(R\'enyi entropies).}

\begin{figure}
\begin{center}
\includegraphics[width=1.1\textwidth]{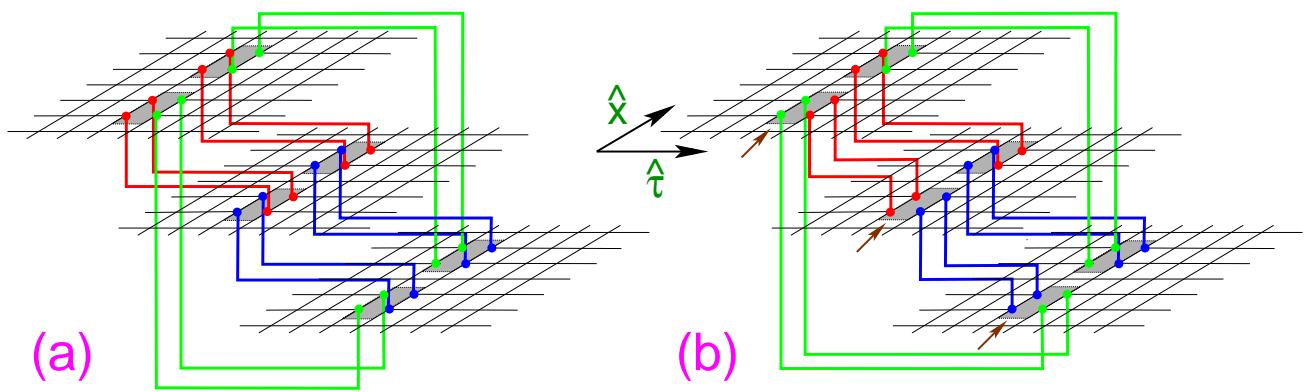}
\end{center}
\caption{ Lattice version of the $n-$sheeted
 Riemann surface ${\mathcal R}_n$ (see ({\bf a})) and 
 the surface (see ({\bf b})) ${\mathcal K}_n$ obtained 
 sewing together $n=3$ independent replicas. On each replica 
 $\hat x$ and $\hat \tau$ stand respectively for the spatial and 
 imaginary time  directions. The shadow is to highlight the position
 of the cut, while colored links 
 connect points on different replicas.
  We show the case of two disjoint intervals
 of lengths $\ell_1=\ell_2=2$ at distance $d=1$. In 
 ({\bf a}) ({\bf b}) in each replica (plane) periodic boundary 
 conditions on both sides are assumed and the partial transposition 
 is done with respect to the degrees of freedom living on the cuts 
 in front (marked by the arrows in (b)). 
}
\label{cartoon_geom}
\end{figure}

We recall that the moments of the reduced density matrix $\Tr\rho_A^n$ 
can be obtained  considering  the Euclidean partition function 
over the so called  $n$-sheeted Riemann surface ${\cal R}_n$ 
 (see~\cite{cc-05p,cc-rev}). In order to  define ${\cal R}_n$ we restrict
 for the moment to the case of $A$ being a single interval. Let us start 
with the set of $n$ independent replicas (sheets). 
Given the sheet  (of area $L\times L_\tau$) corresponding 
to each replica, we consider two points of its dual lattice lying along the
 spatial direction (these would correspond to  the endpoints of subsystem A). 
Then we define a ``cut'' $\lambda$  as the straight line joining the two points,
  its length being the length of subsystem $A$.
The $n-$sheeted Riemann surface is defined by assuming that 
all the links starting from points on the $k-$th replica and intersecting 
the cut connect to points on  the  replica  $k+1(\textrm{mod}\, n)$. 

The generalization of ${\cal R}_n$ to multi intervals is done in the 
 straightforward way.
 As an example in Fig.~\ref{cartoon_geom} ({\bf a}) we show the $3-$sheeted
 Riemann surface with $L=L_\tau=7$ and $\lambda$ made of two disjoint
intervals of equal length $\ell=2$ at distance $d=1$. 

We now introduce the coupled action $S^{(n)}$ over the $n-$sheeted Riemann surface 
as
\begin{equation}
\label{coup_act}
\fl
S^{(n)}(\{\phi_k\})=\sum\limits_{k=1}^n\sum_{\langle ij\rangle \nrightarrow
  \lambda}F(\phi_k(i),\phi_k(j))+ \sum\limits_{\langle ij\rangle\rightarrow
  \lambda}F(\phi_k(i),\phi_{k+1 (mod \,n)}(j)) 
\end{equation}

where $\langle ij\rangle\rightarrow\lambda$ ($\langle ij\rangle\nrightarrow\lambda$) 
 denotes the links which (do not) cross
the cut $\lambda$. Defining as $Z_n[\lambda]\equiv \int\prod_k{\cal D}
[\phi_k]e^{-S^{(n)}(\{\phi_k\})}$ the partition 
function obtained from~\eref{coup_act}, finally $\Tr\rho^n_A$ is 
given by~\cite{cc-04} 

\begin{equation}
\label{ratio_ren}
\Tr\rho^n_A=\frac{Z_n[\lambda]}{Z^n}
\end{equation}

which is formula~\eref{ratio_0} introduced in section~\ref{summ_cft}.
As a final remark one should stress that, in order to recover $\Tr\rho_A^n$ for 
the one dimensional quantum system,  the limit $L_\tau=\beta\to\infty$ has to be taken. 
In actual Monte Carlo simulations, however, it is sufficient to consider very 
elongated lattices with $L\ll L_\tau$~\footnote{We checked that in our simulations
 for both the Ising model and XY model the choice $L_\tau/L\sim 10$ was enough to ensure
 $\beta=\infty$ within the statistical error bar (cf. also~\cite{cg-08})}.

\subsubsection{Replica representation for the  moments of 
$\rho_A^{T_2}$.}

A representation similar to~\eref{ratio_ren} can be obtained for 
 the moments $\Tr(\rho_A^{T_2})^n$. To this purpose we  consider 
a slight modification of action~\eref{coup_act}. 
Now one has $A=A_1\cup A_2$, the partial transposition being  
done with respect to the degrees of freedom of subsystem $A_2$.  
The  cut $\lambda$ is made of two segments as $\lambda=\lambda_1\cup\lambda_2$, with
 $\lambda_1,\lambda_2$ referring respectively to block $A_1$ and $A_2$. 
The modified action reads 

\begin{eqnarray}
\label{coup_act_neg}
\fl
S^{(n,T_2)}(\{\phi_k\})=\sum\limits_{k=1}^n\sum\limits_{\langle ij
\rangle\nrightarrow  \lambda_1\cup\lambda_2}F(\phi_k(i),\phi_k(j))+ \\
\nonumber \sum\limits_{\langle ij\rangle\rightarrow  
\lambda_1}F(\phi_k(i),\phi_{k+1 ({mod}\,n)}(j)) 
+ \sum\limits_{\langle ij\rangle\rightarrow \lambda_2}F(\phi_k(i),\phi_{k-1 
({mod}\,n)}(j))
\end{eqnarray}

Notice that in~\eref{coup_act_neg} the links crossing 
$\lambda_1$ and $\lambda_2$ connect fields living respectively on 
the replicas $k,k+1(\textrm{mod} \, n)$ and 
$k,k-1(\textrm{mod}\, n)$, which can be seen as the net effect of the
 partial transposition. The geometric object (that we call ${\cal K}_n$) 
over which~\eref{coup_act_neg} is defined  is not the $n-$sheeted Riemann 
surface, but  in general has a different topology. 
 For the case $n=3$ and two intervals
 of length $\ell_1=\ell_2=2$ at distance $d=1$ this is depicted
 in Fig.~\ref{cartoon_geom} ({\bf b}).

After defining  the partition function obtained from the action
~\eref{coup_act_neg} as $Z^{T_2}_n[\lambda]$   one 
has~\cite{cct-12,cct-J-12}

\begin{equation}
\label{ratio_neg}
\Tr(\rho_A^{T_2})^n=\frac{Z^{T_2}_n[\lambda]}{Z^n}
\end{equation}
which is the analog of~\eref{ratio_ren}. Both~\eref{ratio_ren} 
and~\eref{ratio_neg} can be calculated efficiently in Monte 
Carlo simulations.

\subsection{Measuring  the moments of $\rho_A^{T_2}$ 
in Monte Carlo simulations}
\label{MC_observable}

In this section we show how to calculate in Monte Carlo simulations 
the ratio of partition functions~\eref{ratio_neg} 
(with minor changes the same strategy applies to~\eref{ratio_ren}).
We start with defining the operator

\begin{equation}
\label{obs}
{\mathcal O}\equiv \exp\left[-S_\lambda^{(n,T_2)}(\phi_1,\phi_2,\dots,
\phi_n)+\sum_kS_\lambda(\phi_k)\right]
\end{equation}

where $S_\lambda$ and $S^{(n,T_2)}_\lambda$ are the ``cut-linked'' 
actions obtained by considering respectively in~\eref{coup_act}~\eref{coup_act_neg}
only the terms with $\langle ij\rangle$ crossing the cut $\lambda$. 
Then, by definition one has

\begin{equation}
\label{obs_raw}
\langle {\mathcal O}\rangle=\Tr(\rho_A^{T_2})^n=
\frac{Z^{T_2}_n[\lambda]}{Z^n}
\end{equation}

where $\langle\cdot\rangle$ stands for the Monte Carlo average 
 over the fields configurations $\{\phi_k(\vec x,\tau)\}$.  
 These are sampled in the Monte Carlo according to the Boltzmann weights $e^{-\sum_kS(\phi_k)}$  
 constructed from the uncoupled action~\eref{act}. 
 
 At this point an important 
remark is in order: although the direct implementation of~\eref{obs} in  
 simulations is  legitimate, its typical Monte Carlo history shows a 
huge variance, due to the presence of the exponential in the definition of 
 ${\cal O}$. This makes~\eref{obs} not useful in practice.

\subsubsection{The increment trick.}
A better behaving observable, which allows to overcome this issue, is  
 obtained  splitting the cut  $\lambda$ in $s$ smaller parts 
$\lambda^{(i)}$ such that $\lambda
\equiv\bigcup_{i=1}^{s}\lambda^{(i)}$ and $s$ is chosen arbitrarily. 
Defining $\hat \lambda^{(i)}\equiv
\bigcup_{k=1}^i \lambda^{(k)}$ one has the trivial identity

\begin{equation}
\label{prod}
\frac{Z_n^{T_2}[\lambda]}{Z^n}=\prod\limits_{i=0}^s\frac{Z_n^{T_2}
[\hat \lambda^{(i+1)}]}{Z_n^{T_2}[\hat \lambda^{(i)}]}
\end{equation}

For each term in the product in~\eref{prod} one can now write

\begin{equation}
\label{expect}
\langle\widetilde{\mathcal O}\rangle_{\lambda'}\equiv\frac{Z_n^{T_2}[\hat 
\lambda^{(i+1)}]}{Z_n^{T_2}[\hat \lambda^{(i)}]}
\end{equation}

where the modified observable $\widetilde{\mathcal O}$ is defined as

\begin{equation}
\label{obs_mod}
\widetilde{\mathcal O}\equiv \exp\left[S_{\lambda'}^{(n,T_2)}
-S^{(n,T_2)}_{\lambda''}\right]
\end{equation}

with $\lambda''\equiv\hat \lambda^{(i+1)}$ and $\lambda'\equiv 
\hat \lambda^{(i)}$. The Monte Carlo average in~\eref{expect} is 
taken with respect to  the action $S^{(n,T_2)}$ with cut $\lambda'$. 
 Now~\eref{obs_mod} receives contribution only from ``fluctuations'' living on a 
portion of the cut and its variance, compared to the one of~\eref{obs},
 is strongly reduced. 

Further improvements are possible if the  model considered (defined by 
 the Euclidean action $S$)  admits a representation in terms of clusters~\cite{
fort-kast-72,chayes-machta-98} (\`a la Fortuin-Kasteleyn) and can 
be simulated  using a Swendsen-Wang like 
algorithm~\cite{swen-wan-87}. Then it is convenient to express~\eref{obs_mod} 
in terms of cluster-related quantities. It turns out that since clusters 
are non local objects  this improves dramatically the efficiency of 
the procedure highlighted so far.

\subsection{Improved Monte Carlo scheme via the ``cut-linked'' cluster 
representation}
\label{MC_proc_clusters}

In this section we describe the improved Monte Carlo method
to simulate $\Tr(\rho_A^{T_2})^n$ (and $\Tr\rho_A^n$)
 for models that admit a representation in terms of 
 clusters. We restrict to the case of  
2D square lattices, although the method can be extended to 
models defined on arbitrary graphs and any dimension in a 
straightforward way. We first introduce the so called 
{\it random cluster model}~\cite{fort-kast-72}.

Let us consider a square lattice and denote with $e$ a generic
 edge (or link) connecting two nearest-neighbor 
sites $x,y$. Let us also define $E$ as the set of 
all the  links on the lattice and  for each $e\in E$ 
consider a ``link function'' $\omega(e)$ such that $\omega(e)\to\{0,1\}$. 
We  call a link $e$  active (inactive) if 
$\omega(e)=1$ ($\omega(e)=0$), 
while the set of all the possible link configurations is 
 $\Omega$  (i.e. $\Omega\equiv\{0,1\}^E$). 
Given an element $\gamma\in\Omega$ we also
consider the set of the active links 
 ${\cal C}_\gamma\equiv\{e\in E:\omega(e)=1\}$.
Finally, the clusters in the configuration $\gamma$ are defined as 
the connected components of ${\cal C}_\gamma$.

The random cluster model, which depends on the two parameters $p$ 
(probability of activating a link) and $q$ (cluster weight), is 
defined through the partition function 

\begin{equation}
\label{rc_pf}
Z=\sum\limits_{\gamma\in\Omega}\left\{\prod\limits_{e\in E}
p^{\omega(e)}(1-p)^{1-\omega(e)}\right\}q^{k(\gamma)}
\end{equation}

where we denote with $k(\gamma)$ the ``counting function'' giving the
 total number of clusters in  $\gamma$.
Many models in statistical mechanics can be mapped to the random 
cluster model (Ising, Potts models, percolation models are just few 
 well known examples). For instance~\eref{rc_pf} becomes the partition 
function of  the 2D Ising model if one chooses

\begin{equation}
q=2\quad p=1-e^{-\beta J}
\end{equation} 

with $\beta$ as usual the inverse temperature. 

Clearly, the definition of the random cluster model can be extended 
 to the surface ${\cal K}_n$ (or to the $n$-sheeted Riemann surface ${\cal R}_n$) 
in a straightforward way. To this purpose we first observe that the  set of 
all the possible links configurations  does not depend on the presence of the 
cut $\lambda$ (since the total number of links is not affected by  
 the geometry of the surface) 
and is   given  as $\Omega_{{\cal K}_n}=\Omega^n=\{0,1\}^{nE}$. Thus  
the partition function of the random cluster model on 
${\cal K}_n$ (i.e. $Z_n^{T_2}[\lambda]$) reads

\begin{equation}
\label{this}
Z_n^{T_2}[\lambda]=\sum\limits_{\gamma\in\Omega^n}
\left\{\prod\limits_{e\in E}p^{\omega(e)}(1-p)^{1-
\omega(e)}\right\}q^{k_\lambda(\gamma)}
\end{equation}

where $k_\lambda(\gamma)$ is the same counting function as in~\eref{rc_pf}.
 The subscript $\lambda$ is to stress that for a given  
$\gamma$ the number of clusters depends on the cut (and hence on the geometry of the surface), 
i.e. it can be $k_\lambda(\gamma)\ne k_{\lambda'}(\gamma)$ if $\lambda\ne\lambda'$. 

In order to derive the (``improved'') cluster version of~\eref{obs_mod} 
 we observe that the partition function of the random cluster model on the 
surface ${\cal K}_n$ with a different cut $\lambda'$ formally can be obtained 
from~\eref{this} (with cut $\lambda$) as 

\begin{equation}
\label{map}
Z_n^{T_2}[\lambda']=\sum\limits_{\gamma\in\Omega^n}\left\{\prod\limits_{e\in E}
p^{\omega(e)}(1-p)^{1-\omega(e)}\right\}q^{k_\lambda(\gamma)-k_\lambda(\gamma)
+k_{\lambda'}(\gamma)}
\end{equation}

In writing~\eref{map}  we  used that, for each fixed $\gamma$, 
 the term in the curly brackets does not depend on the cut.
Dividing~\eref{map} by $Z_n^{T_2}[\lambda]$ we obtain 
the elegant result

\begin{equation}
\label{obs_clust}
\frac{Z_n^{T_2}[\lambda']}{Z_n^{T_2}[\lambda]}=
\langle q^{-k_\lambda+k_{\lambda'}}\rangle_\lambda
\end{equation}

where the average $\langle\cdot\rangle_\lambda$ is done (as the 
subscript stresses) using the action defined on the 
surface ${\cal K}_n$  with cut $\lambda$. Finally, the cluster 
representation of~\eref{obs_mod} is given by

\begin{equation}
\widetilde{\mathcal O}=q^{k_{\lambda'}-k_{\lambda}}
\end{equation}

 with the choice $\lambda=\hat\lambda^{(i)}$ and $\lambda'=\hat\lambda^{(i+1)}$.
 An important property, which is useful in  
Monte Carlo simulations to speed up the evaluation of the average in~\eref{obs_clust}, 
is that the difference $k_{\lambda'}-k_\lambda$ depends 
\emph{only} on the clusters ``intersecting'' the portion of the cut 
$\lambda\cup\lambda'-\lambda\cap \lambda'$  (i.e. only on the ``cut-linked'' clusters), while 
 other contributions  trivially cancel.

\section{Monte Carlo results: Ising universality class}
\label{MC_result_Is}

In this section we numerically investigate the properties of the 
 scale invariant ratios  $r_n$ and $R_n$ 
in the 1D quantum Ising universality class. The Monte Carlo 
data that we present have been  obtained using the results 
 in section~\ref{MC_procedure}, i.e.   simulating the 2D 
Ising model at the critical point $\beta_c\equiv1/T_c=
1/2\log(1+\sqrt{2})$. In particular we used the improved 
scheme described in~\ref{MC_proc_clusters} (details about the 
simulations can be found in~\ref{appendix}).

\subsection{Two adjacent intervals: the ratios $r_n$}

\begin{figure}
\begin{center}
\includegraphics[width=1.1\textwidth]{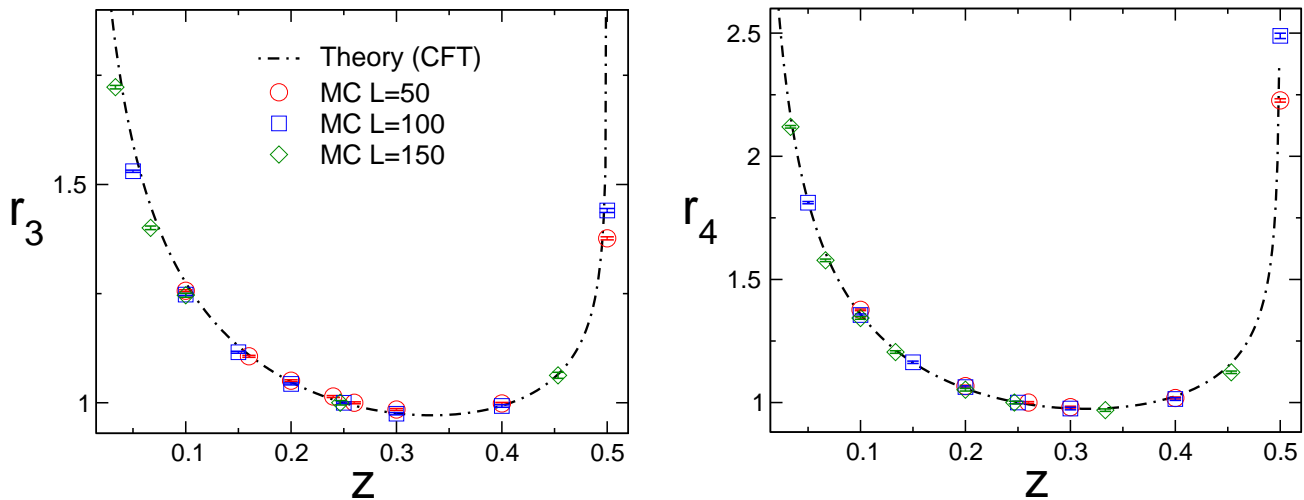}
\end{center}
\caption{ The ratios $r_3$ ({\bf left}) and $r_4$ ({\bf right})
 versus $z\equiv \ell/L$ in the 2D classical critical Ising model.
 Monte Carlo data for several values of $L=50,100,
 150$. The dashed dotted line is the CFT 
 prediction (no fitting parameters). 
 For all the data Monte Carlo errors are smaller than the symbols.
}
\label{fig1}
\end{figure}

Let us  consider two adjacent intervals $A_1,A_2$ 
of equal length $\ell$ (this corresponds to the situation 
shown in  Fig.~\ref{cartoon_0} ({\bf b})). 
In Fig.~\ref{fig1} we show Monte Carlo data for  $r_3$ $r_4$ 
as a function of $z\equiv \ell/L$. We considered in our 
simulations only $L=50,100,150$ and $1\le\ell\le L/2$. 
The asymptotic behavior $r_n(z)$  has been obtained analytically 
in CFT for any $n$ in Ref.~\cite{cct-12}.  This is given in terms 
of~\eref{trA_tran_twist} (cf. section~\ref{summ_cft}) after  
  replacing $\ell\to L/\pi\sin(\pi\ell/L)$. The result is 
reported in Fig.~\ref{fig1} with the dashed-dotted line.

Monte Carlo data 
 show data collapse  for both $r_3,r_4$ and for all the values 
of $L,\ell$ simulated, supporting scale invariance (although this 
is expected only in the asymptotic limit $L,\ell\to\infty$). 
Also, the  behavior of the data is perfectly reproduced by 
the CFT result. Deviations from the  theory are only visible in the 
regions $z\sim 0$ and $z\sim 1/2$. These are understood as 
finite (interval) size effects. Indeed for small $z$ very 
large system sizes are needed to reach the asymptotic  
limit  where CFT  holds. For instance,  $z=0.05$ and $L=150$ 
(which is the largest system size we simulated) corresponds 
to block size $\ell=zL=7.5$, which is apparently too 
small and far from the scaling limit. Similar corrections have 
been observed for $r_3,r_4$ in free bosonic systems (harmonic 
chain)~\cite{cct-12,cct-J-12} (cf. also~\cite{cctt-13} for recent 
results in the Ising chain using TTN techniques).

\subsection{Two disjoint intervals: the ratio $R_3$}

We now discuss the case of two disjoint intervals 
(geometry in Fig.~\ref{cartoon_0} ({\bf c})) focusing on the behavior 
of the ratios $R_n(y)$ (we restrict to $n=3$). We first remind that 
for any $n$, in the limit $L,\ell\to\infty$, $R_n(y)$ are universal functions 
of the harmonic ratio $y$, and  in a CFT  are given 
by~\eref{Rn_cft} (cf. section~\ref{summ_cft_Rn_Is} for the analytical results
 in the Ising case).  

\begin{figure}[t]
\begin{center}
\includegraphics[width=.8\textwidth]{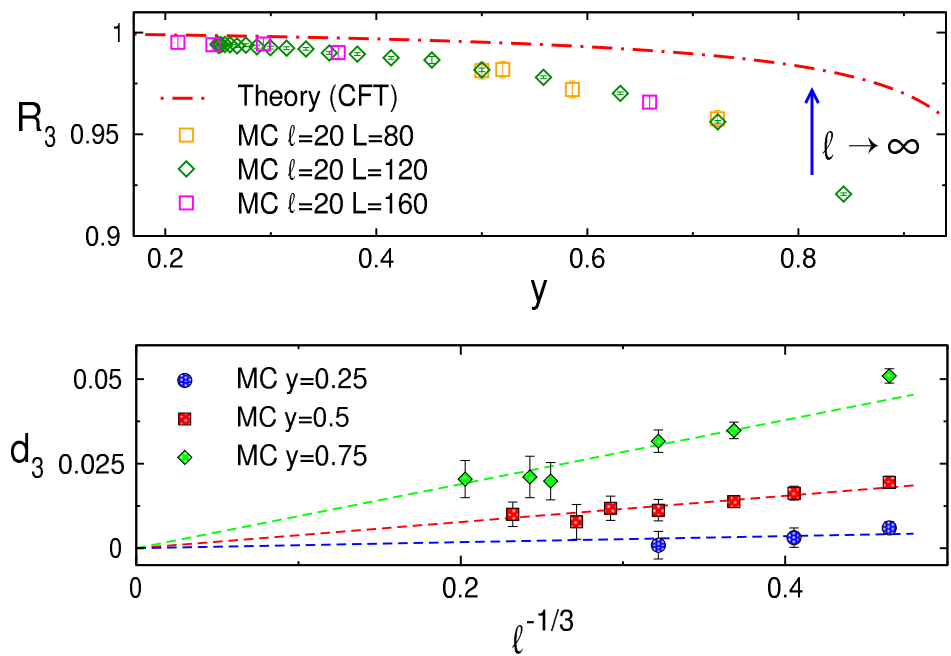}
\end{center}
\caption{ Unusual scaling corrections for the  ratio $R_3(y)$ in the Ising 
 universality class. 
 ({\bf top}) $R_3(y)$ as a function of the harmonic ratio $y$ (Monte 
 Carlo data for two disjoint intervals of length $\ell= 20$ 
 and total system sizes $L=80,120,160$). 
 Different values of $y$ are obtained varying the distance 
 between the two intervals.
 The CFT prediction (dashed dotted line) is recovered in the asymptotic 
 limit (i.e. $\ell \to\infty$), as stressed by the vertical arrow. 
 ({\bf bottom}) Plot of $d_3(y) \equiv R^{CFT}_3(y)-R_3 (y)$  
 versus $\ell^{-1/3}$. Monte Carlo data at fixed $y = 0.25, 0.5, 0.75$ and 
 $10\le\ell\le120$. The dashed lines is the fit to $d_3=a_0\ell^{-1/3}$ 
 with $a_0$ the only fitting parameter. In both plots the Monte Carlo 
 statistical error bar is often smaller than the symbols.
}
\label{fig3}
\end{figure}

In Fig.~\ref{fig3} ({\bf top}) we show Monte Carlo data for $R_3(y)$ 
for several values of the harmonic ratio $0\le y\le 1$ and 
$L=80,120,160$ (data at fixed $\ell=20$). Data at different $y$ are 
obtained varying the distance between the two intervals 
(according to formula~\eref{cross_ratio}). 

Remarkably, all the data 
for different sizes $L$ collapse on the same curve within the Monte Carlo 
statistical error,  meaning that  finite $L$  scaling 
corrections are not visible. Nonetheless, at finite $\ell=20$ 
Monte Carlo data do not match the theoretical (CFT) curve 
(dashed-dotted line), suggesting that finite $\ell$ corrections 
are present. The   CFT result is only recovered in the limit 
$\ell\to\infty$. General renormalization group arguments 
suggest the behavior

\begin{equation}
\label{scal_corr}
R_n(y)=R^{CFT}_n(y)+\ell^{-\omega_n}a_n(y)+\cdots
\end{equation}

with $R_n^{CFT}(y)$ the asymptotic scaling function given by
formula~\eref{Rn_cft}, while $\omega_n$ and $a_n(y)$ are 
respectively the exponent and amplitude of the scaling 
corrections. The dots in~\eref{scal_corr} denote more irrelevant 
terms. Note in~\eref{scal_corr} the dependence of $a_n$ on the 
 harmonic ratio $y$. The very same behavior  (upon replacing $R_n(y)\to 
{\cal F}_n(y)$ in~\eref{scal_corr} and fixing $\omega_n=1/n$) 
is shown by the scaling corrections of the mutual information~\cite{
fps-08,cg-08,cct-09,ch-04,ffip-08,atc-09,atc-11,ip-09,fc-10,
fc-10b,c-10,cct-11,f-12}. 

\begin{figure}[t]
\begin{center}
\includegraphics[width=.75\textwidth]{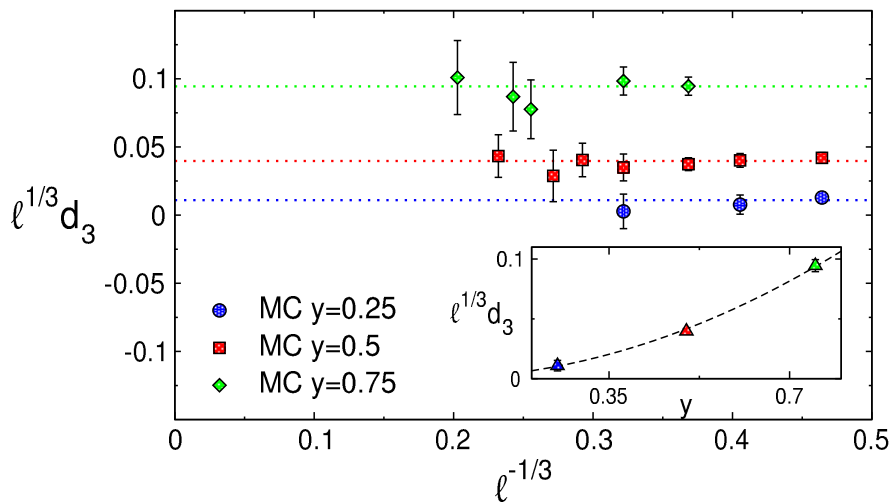}
\end{center}
\caption{ Amplitude $a_3(y)$ of the unusual corrections extracted
 as $a_3(y)=\ell^{1/3}d_3(y)$ with $d_3$ defined as in Fig.~\ref{fig3}.
 Same Monte Carlo data as in Fig.~\ref{fig3} at fixed $y=1/4,1/2,3/4$. 
 We plot $a_3\equiv \ell^{1/3}d_3(y)$ versus $\ell^{-1/3}$. The 
 dotted lines are fits to a constant. Inset: fitted values for $a_3$ 
 (dotted lines in the main Figure) plotted versus $y$. The dashed 
 line is a fit to $Ay^2$.
}
\label{fig3x}
\end{figure}

More generally, entanglement related quantities are 
known to exhibit unusual scaling corrections. These are induced 
by the presence of the conical singularities (at the edges of 
 the subsystem) needed to write the reduced density matrix 
(and the partially transposed one) in the quantum field theory 
language~\cite{cc-10}. 

At difference with usual corrections, arising due to  the presence of 
irrelevant (in the renormalization group sense) operators in the theory, 
 the unusual ones are induced by a {\it local} insertion, near the conical singularity, of an 
 operator that would be {\it relevant} in the bulk. 
  The scaling dimension $x$ of the  operator determines the exponent
 of the unusual corrections as $\omega_n=2x/n$.
 For this reason the analysis of  entanglement corrections 
provides a tool to unveil universal information about critical systems. 

On the other hand, it is a hard task in general to identify the 
relevant operator inducing  the unusual corrections because  
 any operator which does not break the symmetries of the model is 
 in principle allowed~\cite{cc-10}. For the Ising universality 
class  it has been shown  that this is  the Majorana operator 
(with $x=1/2$), implying $\omega_n=1/n$.  
It is natural to expect that the same one induces  
the scaling corrections of $R_n(y)$, although their exponent 
 could be different (i.e. $\omega_n\ne1/n$). 

To clarify this issue in Fig.~\ref{fig3} 
({\bf bottom}) we show $d_3(y)\equiv R_3^{CFT}(y)-R_3(y)$ at fixed 
$y=1/4,1/2,3/4$ and $\ell$ in the range $10\le\ell\le 120$. Clearly, 
Monte Carlo data support the behavior as $\ell^{-1/3}$. We also mention
that a fit to $a/\ell^{b}$ leaving $b$ as a free parameter 
gives the exponent $b=0.4(1)$ again consistent with $b=1/3$. 
This allows to  conclude that in the critical Ising spin 
chain the unusual corrections for $R_n(y)$ are 
 of the same form as the ones for the mutual information.

We provide complementary information about the scaling corrections
 in Fig.~\ref{fig3x}, plotting their amplitude $a_3(y)$ as a function 
of $\ell^{-1/3}$. Here $a_3(y)$ is extracted from the Monte Carlo data 
shown in Fig.~\ref{fig3} as $a_3(y)=\ell^{1/3}d_3(y)$.
Its behavior confirms the correctness of the 
scaling corrections exponent $\omega_3=1/3$. A precise estimate
 for $a_3(y)$ is obtained by  fitting the data with a constant 
(dotted lines in the Figure). The results are shown in the  
 inset as a function of the harmonic ratio $y$ and are well described  
 by a parabolic behavior as $\sim y^2$  up to $y=3/4$ (dashed line 
 in Fig.~\ref{fig3x}: Inset). One should stress 
that this is different from what observed for the mutual information. 
For instance, it has been shown in Ref.~\cite{atc-09} that 
the amplitude of the corrections to ${\cal F}_2(y)$ is very well  
described by $y^{1/4}$, pointing to a much slower decay
  in the limit $y\to0$.

\section{ Monte Carlo results: 2D XY model at the 
 BKT transition}

In this section we numerically investigate the behavior 
of the ratios $r_n$ and $R_n$ for the free compactified 
boson theory with compactification radius 
$r_{circle}=\sqrt{2}$  (equivalently $\eta=4$ or, in 
terms of the Luttinger parameter, $K_L=2$)~\cite{ginsparg-89}. 
This describes (a part from logarithmic corrections) 
the critical properties of the 2D classical XY model at the 
BKT phase transition or, equivalently,  the long wavelength
behavior of the spin-$\frac{1}{2}$ quantum Heisenberg XXZ chain 
at anisotropy $\Delta=-1/\sqrt{2}$.

The Monte Carlo data we present have been  obtained using 
 the method outlined in~\ref{MC_procedure}, simulating  the
2D XY model at $\beta_{BKT}$ (for more details about the 
 simulations cf.~\ref{appendix}). Since for the XY model there 
is no efficient implementation of the Swendsen-Wang algorithm, we 
could not  use the improved scheme provided in section~\ref{MC_proc_clusters}.

Nonetheless, one general result of this section is 
that the scheme described in section~\ref{MC_procedure}
 is effective for models  with {\it continuous} 
degrees of freedom. This is verified in a preliminary step of our 
analysis (cf. section~\ref{ccXY})  by calculating 
$\Tr\rho_A^n$ ($n=2,3,4$ with $A$ a single block) and
 checking its scaling behavior against the well
 known CFT result~\eref{single}. 

We then focus (section~\ref{rRnXY}) on the ratios $r_3,r_4$ 
finding that, already for finite 
intervals, their behavior is well reproduced by the CFT results 
in Ref.~\cite{cct-12,cct-J-12}. 
Surprisingly, this is also the case for the ratio $R_3(y)$ for finite
 (large enough) blocks size, signaling that  corrections 
are smaller  than in the Ising case.

\subsection{ Single interval: the moments of $\rho_A$
and the central charge}
\label{ccXY}

In this section we validate the Monte Carlo procedure 
outlined in section~\ref{MC_procedure} for models with continuous 
degrees of freedom. In particular we discuss Monte Carlo results
for the moments $\Tr\rho_A^n$ ($n=2,3,4$) of the reduced density matrix,  
demonstrating that their scaling behavior is fully reproduced (as expected) by 
the CFT result~\eref{single}.
Besides its relevance as a benchmark of the method, our analysis 
suggests that, even for models defined in terms of continuous variables, 
entanglement based quantities are effective tools to extract 
their  central charge~\footnote{ 
Another available method is based on the behavior 
of the finite size corrections  of  the free energy~\cite{ecorr}, 
which, however, is difficult to obtain accurately in Monte Carlo 
simulations.}.

\begin{figure}[t]
\begin{center}
\includegraphics[width=.85\textwidth]{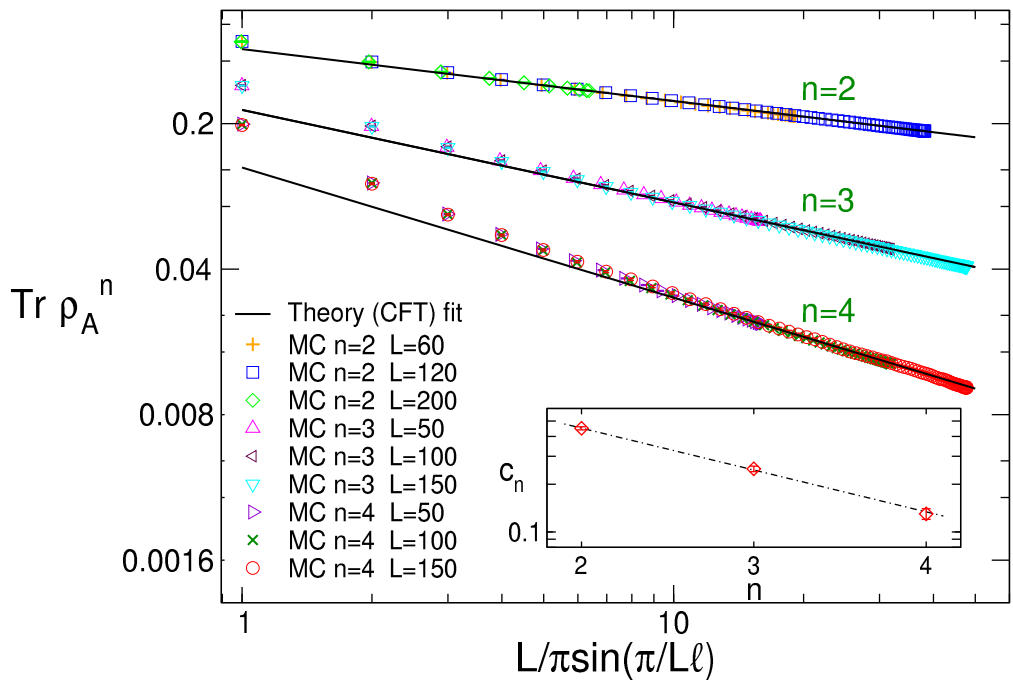}
\end{center}
\caption{ Central charge $c$ of the Berezinskii-Kosterlitz-Thouless 
 universality class (2D XY model at the BKT point $\beta_{BKT}=1.1199(1)$).
 We show $\textrm{Tr}\rho_A^n$ (Monte Carlo data) versus 
 $L/\pi\sin(\pi\ell/L)$ for $n=2,3,4$, several 
 values of $L$ $50\le L\le 200$, and $1\le\ell\le 75$. The Monte Carlo error
 bar is smaller than the size of the symbols. Continuous
  lines are one parameter fits to the asymptotic CFT behavior~\eref{single_fin} 
 after fixing $c=1$. Inset: Fitted values of the 
 non universal constant $c_n$ as a function of $n$ (note the logarithmic
 scale on the $y$-axis). The dashed line is a fit to an exponential
 decay.
 }
\label{XY_single}
\end{figure}

We start with reminding that  the scaling behavior of $\Tr\rho^n_A$ 
in the asymptotic limit  is 
given in CFT as (cf. section~\ref{summ_cft})

\begin{equation}
\label{single_fin}
\textrm{Tr}\rho_A^n\sim c_n\left(\frac{L}{\pi}\sin
\frac{\pi\ell}{L}
\right)^{-\frac{c}{6}(n-\frac{1}{n})}
\end{equation}

and $c=1$ for the XY model. In Fig.~\ref{XY_single} we show Monte Carlo 
data for $\textrm{Tr}\rho_A^n$ with $n=2,3,4$ and $50\le L\le 200$, 
 $1\le\ell\le 75$. For each $n$ and different $L$ all the data 
collapse on a single curve meaning that $L$ dependent scaling 
corrections are not visible within the Monte Carlo error bar.

The continuous (black) lines are given by~\eref{single_fin}
where we fix the central charge $c=1$, fitting the non universal 
constant $c_n$. For $n=2$ the CFT prediction reproduces the 
behavior of the MC data in the whole range $1\le\ell\le 75$.
This is not the case at  $n=3,4$  where larger $\ell$ dependent
 scaling corrections are present and bigger systems are needed
 in order to reach the asymptotic limit. In fact,  
 agreement with MC data is observed only at $L_c\equiv L/\pi\sin(\pi\ell/L)
\gtrsim 10$ and $L_c\gtrsim 20$ for respectively $n=3,4$. The same 
 increasing trend in the scaling corrections of $\Tr\rho_A^n$
 ( upon increasing the R\'enyi index $n$) is expected in a generic system 
 described  by the Luttinger liquid~\cite{ccen-10}.

To have an independent estimate of the central charge we 
performed fits leaving $c$ as a free parameter in~\eref{single_fin}.
For $n=2$ a fit to~\eref{single_fin} (discarding the data up to 
$L_c\lesssim 8$) gives  $c_2=0.45(1)$ $c=0.99(1)$ with $\chi^2/\textrm{DOF}
\approx 1.3$. Similarly for $n=3,4$ one gets respectively $c_3=0.25(1)$ 
$c=1.00(3)$ and $c_4=0.13(1)$ $c=1.00(3)$.
Finally, in Fig.~\ref{XY_single} (inset) we  show the 
fitted value of the non universal constant $c_n$ as a function
of $n$. Clearly $c_n$ shows an exponential decay and
is well described by the function $Ae^{-n/n_0}$
with $A\approx 1.52$ and $n_0\approx 1.64$. 
Similar exponential behavior of $c_n$ has been already observed in the 
1D XXZ spin chain~\cite{ccen-10}.

\subsection{The ratios $r_n$, $R_n$}
\label{rRnXY}

\begin{figure}[t]
\begin{center}
\includegraphics[width=1.1\textwidth]{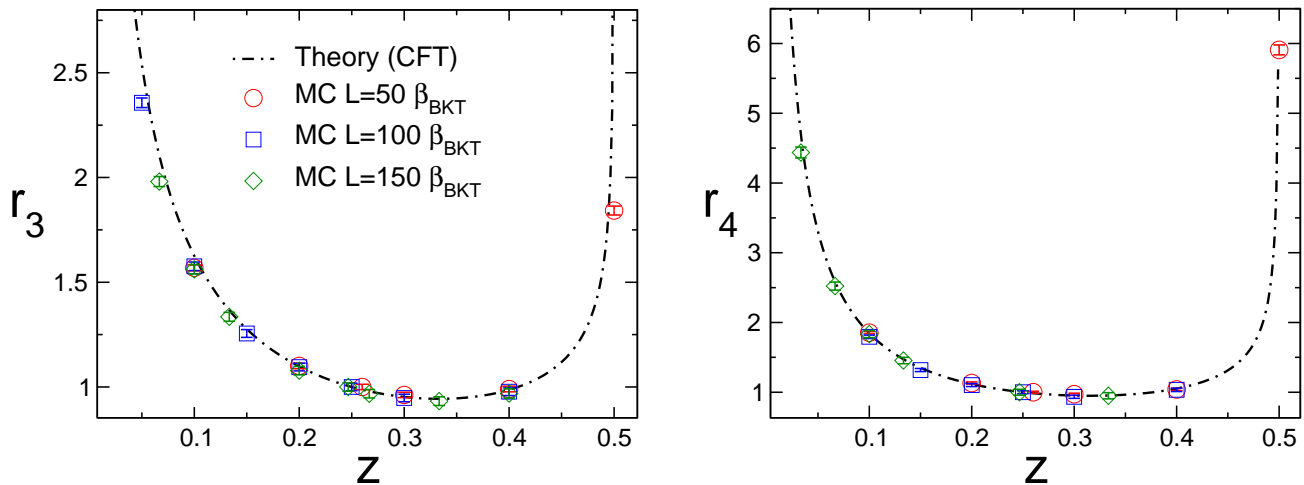}
\end{center}
\caption{ The ratios $r_3$ ({\bf left}) and $r_4$ ({\bf right})
 versus $z\equiv \ell/L$ in the 2D XY model at the BKT phase
 transition. We show Monte Carlo data for several values 
 of $L=50,100,150$. The dashed dotted line is the CFT prediction
 obtained from~\eref{rn_cft} (no fitting parameters). 
 For both $r_3,r_4$ the error bar is  present although 
 smaller than the symbols.
}
\label{fig2}
\end{figure}

We now proceed to discuss the behavior of the ratios $r_n$.
In Fig.~\ref{fig2} we show Monte Carlo data for $r_3,r_4$ 
versus $z\equiv \ell/L$ and system sizes 
$L=50,100,150$. We considered $\ell$ in the range $1\le\ell\le 75$.
As for the Ising universality class  Monte
Carlo data for different values of $L$ are on the same curve 
(scale invariance) which is remarkably well described (no fitting parameters) 
 by the CFT result (cf.~\eref{rn_cft})  (dashed-dotted line in the Figure). 
Similarly to the Ising case finite size deviations are visible  
in the regions $z\sim 0, 1/2$. Notice also that,  
 due to the larger value of the  central charge, both $r_3$ $r_4$ range 
in a larger interval, as functions of $z$, compared to Fig.~\ref{fig1}.  

\begin{figure}[t]
\begin{center}
\includegraphics[width=.85\textwidth]{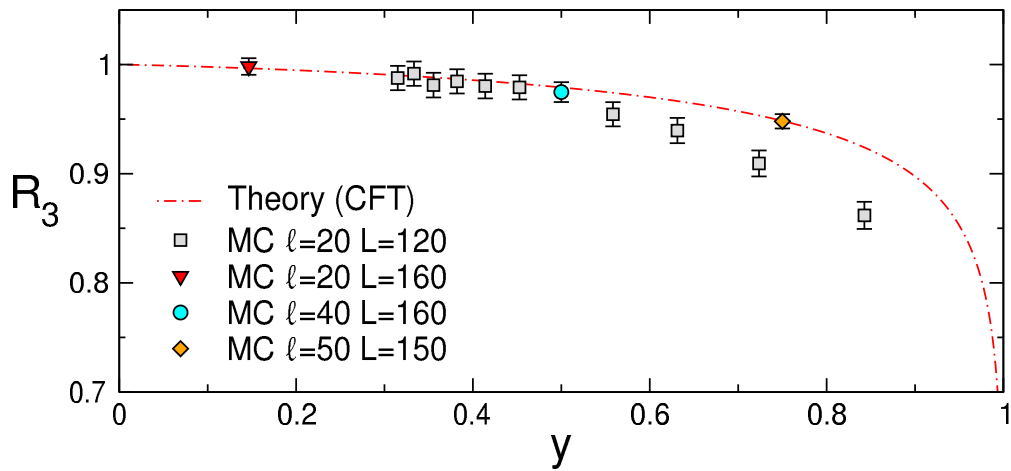}
\end{center}
\caption{ Scale invariant ratio $R_3(y)$ as a function of 
 the harmonic ratio $y$ (Monte Carlo data for the 2D XY model at 
 the BKT phase transition and $\ell=20,40,50$). 
 The dashed dotted line is the CFT prediction from~\eref{fn_free_c}.
 At fixed  $\ell=20,L=120$ different values of $y$  
 correspond to different distances between the two blocks. 
}
\label{fig3_xy}
\end{figure}

The behavior of the universal ratio $R_3(y)$ is more surprising.
 Monte Carlo data for $R_3(y)$ as a function of the
harmonic ratio $y$ are reported in Fig.~\ref{fig3_xy} 
 (data for $\ell=20,40,50$). Focusing on the configuration  
 with $\ell=20,L=120$  Monte Carlo data are
 in reasonable agreement with the CFT curve (dashed-dotted line)
 for $y\lesssim 1/2$. Deviations from the theory are visible at $y>1/2$ and 
increase as a function of $y$, as observed in Ising model 
(cf. Fig.~\ref{fig3}). One striking difference, however, is that 
 apparently they  decay faster upon increasing the block size $\ell$,
 which could suggest a large value of the corrections exponent
 $\omega_3$. This is evident from the perfect match
 between theory and Monte Carlo for $\ell=40,50$ at respectively
 $y=1/2$ and $y=3/4$. One should mention, however, that 
the precision of the data does not allow to extract a reliable
 estimate of the exponent $\omega_3$ and a more careful analysis
 would be needed to reach a conclusion.


\section{Conclusions}

In this work we studied the moments $\Tr(\rho_A^{T_2})^n$ 
of the partially transposed reduced density matrix $\rho_A^{T_2}$
 in critical quantum spin chains. Given a bipartition of the chains 
into two parts $A$ and $B$ with $A$ being made of two equal blocks  as 
$A=A_1\cup A_2$,  $\rho_A^{T_2}$ is obtained from the reduced density
 matrix $\rho_A$ by performing the partial transposition with respect 
to the degrees of freedom of  $A_2$.
We considered both the situations with two adjacent and disjoint blocks, 
 defining, respectively for the two cases, the ratios
 $r_n$ and $R_n(y)$, which are scale invariant at the critical point.
Using  a new numerical method based on Monte Carlo simulations 
we  characterized their scaling  behavior, in the  critical Ising spin 
chain and  the 1D anisotropic Heisenberg   XXZ model at 
$\Delta=-1/\sqrt{2}$. 

The  long wavelength properties of both models are described by 
well known conformal field theories: the first 
corresponds in the continuum to  a free Majorana fermion (central 
charge $c=1/2$), while the second is  the free compactified boson 
at compactification radius  $r_{circle}=\sqrt{2}$ ($c=1$). 
In $d+1=2$ dimensions these  also correspond respectively to the 2D 
classical critical Ising model  (Ising  universality class) and  the 2D 
classical critical XY model (Berezinskii-Kosterlitz-Thouless universality 
class). 

The results of our work can be summarized as follows:

\begin{itemize}
\item [({\it i})] Exploiting the mapping between a quantum system in $d$ dimensions 
 and a classical one in $d+1$ we developed a new numerical scheme 
 to calculate all the moments of the partially transposed reduced 
 density matrix $\rho_A^{T_2}$ in classical Monte Carlo simulations. 
 The method generalizes the one used 
 in~\cite{cg-08,atc-09,atc-11} for calculating the moments of 
 the reduced density matrix $\rho_A$ and is effective for systems with 
 both discrete (Ising model) and continuous (XY model) degrees of freedom. 
 For models admitting a representation in terms of cluster variables
  we provided a  modified (improved) version of the algorithm.
\item [({\it ii})] For two adjacent blocks $A_1,A_2$ we studied 
 the behavior (as a function of the length of the two blocks)
  of the ratio $r_n$ (with $n=3,4$) in both the critical 
 Ising quantum spin chain and the 1D anisotropic Heisenberg XXZ model 
 at  $\Delta=-1/\sqrt{2}$. In both cases we numerically 
 demonstrated (for the first time in non free models~\footnote{See 
 also~\cite{cctt-13} for a recent independent check in the Ising spin chain 
  using TTN techniques.}) 
 that their behavior is well described  by CFT (results in  
 Ref.~\cite{cct-12,cct-J-12}). 
\item [({\it iii})] For two disjoint intervals we studied the scaling 
 properties of $R_3(y)$, which in the asymptotic limit 
 ($\ell\to\infty$) is a universal function of the harmonic ratio $y$.
  For the 1D quantum Ising model $R_3(y)$ exhibits strong (finite size) 
 unusual scaling corrections~\cite{cc-10}.  
 Their  exponent $\omega_3=1/3$  is the same as that found
 for the mutual information between the two blocks~\cite{atc-09}.  After 
 taking into account scaling corrections Monte Carlo data  show full agreement 
 with the asymptotic (i.e. for infinite blocks) CFT result. For the XXZ chain, 
 although  both usual and unusual corrections are expected, they appear 
 to be much smaller and already for sizes $\ell\sim 50$ we find that $R_3(y)$ is 
 in excellent agreement with the theory. 
\end{itemize}

\paragraph*{ {\bf Acknowledgments}:}

I would like to thank Pasquale Calabrese for drawing to my attention this
 problem. I would also like to thank Andreas L\"auchli and Maurizio Fagotti 
 for collaboration in a related project and interesting discussions. I thank
the authors of Ref.~\cite{cctt-13} for sharing, after this work was completed,
their results before publication.  
This work was partly done during the workshop ``New quantum states 
of matter in and out of equilibrium''
 at the Galileo Galilei Institute in Florence whose
hospitality is gratefully acknowledged. Simulations have been performed on the 
local cluster at the Max Planck Institute for the Physics of the Complex
 Systems (MPIPKS) in Dresden.

\appendix
\section{Monte Carlo simulations}
\label{appendix}
\subsection{The algorithm}
In this appendix we give some details about the Monte Carlo algorithms  
used for the simulations.

\subsubsection{Ising model.}
For the simulation of the 2D Ising model we employed the standard implementation 
of the Swendsen-Wang algorithm as described in~\cite{swen-wan-87}. 
To generate the random  numbers needed in the Monte Carlo update we used 
the Mersenne twister random number generator~\cite{mersenne} (the same generator was
 used for the simulation of the 2D XY model). The cluster labeling step needed  
in the Monte Carlo update was performed using the standard ``ants in the labyrinth''
 algorithm~\cite{de-gennes-76}. 
\subsubsection{XY model.} For the 2D XY model at the BKT point 
 our Monte Carlo algorithm was a local Metropolis update 
 supplemented with some overrelaxation sweeps. The Metropolis update that we used is 
different from the  one usually found in the literature. At each Metropolis step 
 a random vector $\vec n=(\cos\alpha,\sin\alpha)$ was chosen with $\alpha$ uniformly 
distributed in $[0,2\pi]$. Then for each spin $\vec s_x$ at lattice site $x$  the
 update proposal $\vec s_x^{\,\prime}$ was obtained reflecting $\vec s_x$ with 
respect to $\vec n$, i.e.

\begin{equation}
\vec s_x^{\,\prime}=2(\vec n\cdot \vec s_x)\vec n-\vec s_x
\end{equation}

The proposal was accepted with the standard Metropolis probability 
$\textrm{Min}[e^{-(E-E')},1]$ with $E(E')$ the energies of the configurations 
before ad after the update. This procedure allows  to avoid 
the (expensive) evaluation of the exponential function at each site of 
the lattice (which is the case in the standard Metropolis implementation).

To improve the performances of the Metropolis we added some sweeps of 
overrelaxation. The proposal for the new configuration $\vec s_x^{\,\prime}$ 
in the overrelaxation step is obtained by reflecting $\vec s_x$
 with respect to the local field $\vec S=\sum_{y}\vec s_y$ (the sum is over 
the nearest neighbor sites of $x$):

\begin{equation}
\label{overr_step}
\vec s_x^{\,\prime}=2(\vec s_x\cdot \vec S)\vec S-\vec s_x
\end{equation}

Since the new configuration with $\vec s_x^{\,\prime}$  has by definition the 
same energy (as that with $\vec s_x$) , the update step~\eref{overr_step} is 
accepted with probability one.
However, for this reason, overrelaxation is not ergodic and has to be used 
with some other ergodic algorithm. In our case each 
Monte Carlo step was a combination of one Metropolis sweep followed by
 five overrelaxation steps.  The overrelaxation sweeps have the advantage that, 
although the energy  does not change,  the initial and final configurations can 
be very different, reducing dramatically the critical slowing down. 
It can be shown indeed that the autocorrelation time $\tau$ for the 2D XY model
 at the BKT point, using overrelaxation, can be reduced to $\tau=0.15
\xi^{1.2}$~\cite{gupta-88}  (here $\xi$ is the correlation
 length) while the standard Metropolis gives $\tau\sim\xi^2$. Here it is worth
 stressing that any non local update (such as the Wolff algorithm~\cite{wolff-89}) 
outperforms the procedure outlined above\footnote{ In particular, for the 2D XY model 
 at the BKT point in Ref.~\cite{wolff-89}, using the Wolff algorithm, no sign of the 
 critical slowing down was observed.}. On the other hand the update scheme that we used 
 is quite effective for frustrated systems where there is no available  cluster algorithm~\cite{grosse-97}.

\end{document}